\documentstyle[12pt,epsfig]{article}
\newcommand{\sect}[1]{\setcounter{equation}{0}\section{#1}}

\def\rf#1{(\ref{eq:#1})}
\def\lab#1{\label{eq:#1}}
\def\nn{\nonumber \\}
\newcommand{\beano}{\begin{eqnarray*}}
\newcommand{\enano}{\end{eqnarray*}}
\def\bea{\begin{eqnarray}}
\def\ena{\end{eqnarray}}
\def\be{\begin{equation}}
\def\ee{\end{equation}}

\relax

\font\fld=msbm10 at 12 pt
\font\goth=eufm9 at 12 pt
\newcommand{\fl}[1]{\mbox{\fld #1}}     
\newcommand{\go}[1]{\mbox{\goth #1}}    

\def\a{\alpha}
\def\b{\beta}
\def\d{\delta}

\def\g{\gamma}

\def\l{\lambda}
\def\L{\Lambda}
\def\m{\mu}

\def\p{\phi}
\def\P{\Phi}

\def\ra{\rightarrow}

\def\cl{{\cal L}}

\newcommand{\ad}{\mbox{\rm ad}}

\newtheorem{theo}{Theorem}

\def\ZZ{{\fl Z}}

\def\RR{{\fl R}}
\begin{document}
%
\begin{titlepage}
\vspace{-1cm}
\noindent
\hfill{US-FT/1-00}\\
\vspace{0.0cm}
\hfill{hep-th/0002219}\\
\vspace{0.0cm}
\hfill{February 2000}\\
\phantom{bla}
\vfill
\begin{center}
{\large\bf  Massive Symmetric Space sine-Gordon Soliton}\\
\vspace{0.3cm}
{\large\bf  Theories and Perturbed Conformal Field Theory}\\
\end{center}

\vspace{0.3cm}
\begin{center}
Olalla A. Castro Alvaredo and J. Luis Miramontes
\par \vskip .1in \noindent
{\em 
Departamento de F\'\i sica de Part\'\i culas,\\
Facultad de F\'\i sica\\
Universidad de Santiago de Compostela\\
E-15706 Santiago de Compostela, Spain}\\
\vspace{0.1in}
{\tt castro@fpaxp1.usc.es} {\hskip 0.15truecm} and {\hskip 0.15truecm}
{\tt miramont@fpaxp1.usc.es}\\
\par \vskip .1in \noindent
\normalsize
\end{center}
\vspace{.2in}
\begin{abstract}
\vspace{.3 cm}
\small
\par \vskip .1in \noindent

\noindent
The perturbed conformal field theories corresponding to the massive 
Symmetric Space sine-Gordon soliton theories are identified by
calculating the central charge of the unperturbed conformal field
theory and the conformal dimension of the perturbation. They are described by
an action with a positive-definite kinetic term and a real potential term
bounded from below, their equations of motion are non-abelian affine Toda
equations and, moreover, they exhibit a mass gap. The unperturbed CFT
corresponding to the compact symmetric space $G/G_0$
is either the WZNW action for
$G_0$ or the gauged WZNW action for a coset of the form $G_0/U(1)^p$. The
quantum integrability of the theories that describe perturbations of a WZNW
action, named Split models, is established by showing that they have quantum
conserved quantities of spin $+3$ and $-3$. Together with the already
known results for the other massive theories associated with the non-abelian
affine Toda equations, the Homogeneous sine-Gordon theories, this supports
the conjecture that all the massive Symmetric Space sine-Gordon theories
will be quantum integrable and, hence, will admit a factorizable $S$-matrix.
The general features of the soliton spectrum are discussed, and some 
explicit soliton solutions for the Split models are constructed. In general,
the solitons will carry both topological charges and abelian Noether charges.
Moreover, the spectrum is expected to include stable and unstable particles. 



\end{abstract}
\vfill
\end{titlepage}

\sect{Introduction.}
\ \indent\label{intro}

The non-abelian affine Toda (NAAT) equations~\cite{nt,Luiz} are integrable
(multi-component) generalizations of the sine-Gordon equation for a bosonic
field that takes values in a non-abelian Lie group, in contrast with the usual 
Toda field theories where the field takes values in the (abelian) Cartan
subgroup of a Lie group. In~\cite{ntft} (see
also~\cite{ls}), it was found the subset of NAAT equations that can be written as
the classical equations of motion of an action with a positive-definite
kinetic term, a real potential term bounded from below and, moreover, with a
mass-gap in order to make possible an $S$-matrix description. The resulting
theories were named Homogeneous sine-Gordon (HSG) and Symmetric Space
sine-Gordon (SSSG) theories, which are associated with the different compact
Lie groups and compact symmetric spaces, respectively. Actually, the HSG and
SSSG theories of~\cite{ntft} are particular examples of the deformed coset
models constructed by Park in~\cite{park} and the Symmetric Space
sine-Gordon models constructed by Bakas, Park and Shin in~\cite{sssg3},
respectively, where the specific form of the potential makes them exhibit a
mass gap.

There are some general features of these theories that have to be 
emphasized. The first one is that all of them have soliton solutions.
Taking into account that both the HSG and SSSG theories are described by
actions with sensible properties, this is in contrast with the usual affine
Toda field theories where the condition of having soliton solutions
(imaginary coupling constant) always leads to an ill defined
action which makes the quantum theory problematic~\cite{illToda}. The second
is that they are defined by an action of the form
\be
S[h]\> =\> {1\over\beta^2}\> \Bigl\{ S_{WZNW}[h]\> -\> \int d^2x\>
V(h)\Bigr\}\>,
\lab{ActGen}
\ee
where $h=h(x,t)$ is a field that takes values in a non-abelian
compact group $G$, $V(h)$ is some potential function on the group
manifold, and $S_{WZNW}$ is either the WZNW action for the group $G$ or the
gauged WZNW action for a coset of the form $G/H$, where $H$ is an abelian
subgroup of $G$ to be specified. Therefore, if the quantum theory is to be well
defined then the coupling constant has to be quantized: $\beta^2= 1/k$, for some
positive integer $k$ (see~\rf{level} for a more precise form of this
quantization rule). Such a quantization does not occur in the sine-Gordon
theory or the usual affine Toda theories because the field takes values in an
abelian group in those cases. An important consequence of this is that, in the
quantum theory, the $\beta^2$ will not be a continuous coupling constant.
However, the quantum theory will have other continuous coupling constants that
appear in the potential and, in particular, determine the mass spectrum. 
A third feature which follows from~\rf{ActGen} is that these
theories are naturally described as perturbations of conformal field theory
(CFT) coset models. Therefore, they will provide a Lagrangian formulation for
some already known integrable perturbations of CFT's and, furthermore, they will
also lead us to discovering new ones. Finally, these theories are expected to
exhibit realistic properties of quantum particles not captured by other
integrable field theories; for instance, the presence of unstable particles.

The main quantum properties of the HSG theories are quite well understood.
They are integrable perturbations of the $G$-parafermion theories,
which are coset CFT's of the form $G_k/H$, where $G$ is a
compact simple Lie group of rank~$r_g$, $H\simeq U(1)^{r_g}$ is a maximal
torus, and the `level'~$k$ is a positive integer. The perturbation is given by a
spinless primary field of conformal dimension $\Delta =\bar{\Delta}=
h_g^\vee/(k+h_g^\vee)$, with
$h_g^\vee$ the dual Coxeter number of~$G$. The simplest HSG theory is associated
to $G=SU(2)$,  whose equation of motion is the complex sine-Gordon
equation~\cite{park}. This theory corresponds to the perturbation of the usual
$\ZZ_k$-parafermions by the first thermal operator~\cite{CSGBAK}, whose exact
factorizable scattering matrix is the minimal one associated to
$a_{k-1}$~\cite{paras,CSGMat}. The quantum integrability of the HSG theories
for arbitrary~$G$ was established in~\cite{hsg1}, its soliton spectrum was
obtained in~\cite{hsg2}, and a proposal for the exact factorizable 
$S$-matrices of the theories related to simply laced Lie groups $G$ has been
made in~\cite{hsg3}. The main feature of those
$S$-matrices is that they possess resonance poles
which can be associated directly to the presence of  unstable particles in
the spectrum via the classical Lagrangian. Actually, to the best of our
knowledge, these are the only known integrable quantum field theories that
describe unstable particles. The
$S$-matrices of~\cite{hsg3} have been probed in~\cite{TBA} using the
thermodynamic Bethe ansatz. In particular, this analysis confirms the
expected value of the central charge of the unperturbed CFT for any simply
laced~$G$, and supports the interpretation of the resonance poles as a
trace of the existence of unstable particles in the theory. These
scattering matrices have been recently generalized in a Lie algebraic sense
by Fring and Korf in~\cite{Andreas}. 

In contrast, the quantum properties of the generality of SSSG theories are not
known, and the purpose of this paper is to partially fill this gap. Namely, we
will find the class of perturbed CFT's corresponding to the SSSG theories, 
investigate their quantum integrability, and discuss the general features of
their spectrum of solitons.

The SSSG theories are related to a
compact symmetric space $G/G_0$, with a $G_0$-valued field, and describe
perturbations of either the WZNW CFT corresponding to $G_0$ or a coset
CFT of the form $G_0/H$, where $H\simeq U(1)^p$ is a torus of $G_0$, not
necessarily maximal. The equations of motion of this kind of theories for more
general choices of the normal subgroup $H$ were originally considered in the
context of the, so called, reduced two-dimensional $\sigma$-models~\cite{sssg1},
although their Lagrangian formulation was not known until much
later~\cite{sssg3}. The results of~\cite{ntft,ls} show that they fit quite
naturally into the class of non-abelian affine Toda theories and, what is more
important, that the condition of having a mass gap requires that $H$ is either
trivial or abelian.  The simplest SSSG theories are the ubiquitous sine-Gordon
field theory, which corresponds to $G/G_0 = SU(2)/SO(2)$, and the
complex sine-Gordon theory, which is related this time to $Sp(2)/U(2)$~\cite{ls}
(recall that it is also the HSG theory associated to $SU(2)$). Actually, these
two theories serve as paradigms of what can be expected in more complex
situations. Another theories already discussed in the literature that belong to
the class of SSSG theories are the integrable perturbations of the $SU(2)_k$
WZNW model and its $\widetilde{so(2)}$ reduction constructed by
Brazhnikov~\cite{brazhnikov}. Both of them are related to the symmetric space
$SU(3)/SO(3)$ and, moreover, the second is identified with the perturbation of
the usual $\ZZ_k$-parafermions by the second thermal operator.

The classification of the SSSG theories as perturbed CFT's is achieved through
the calculation of the central charge of the unperturbed CFT and
the conformal dimension of the perturbation. Since the unperturbed CFT is always
a coset CFT of the form $G_0/H$, the calculation of its central charge is
straightforward. In contrast, the calculation of the conformal dimension of
the perturbation requires the knowledge of the structure of the symmetric
space. The symmetric space $G/G_0$ is associated with a Lie algebra
decomposition $g= g_0\oplus g_1$ that satisfies the commutation relations
\be
[g_{0}\>,\> g_{0}]\subset g_{0}\>, \quad
[g_{0}\>,\>g_{1}]\subset g_{1}\>, \quad
[g_{1}\>,\>g_{1}]\subset g_{0}\>,
\lab{gradation}
\ee
where $g$ and $g_0$ are the Lie algebras of $G$ and $G_0$, respectively.
Then, the conformal properties of the perturbation depend on the
structure of the representation of $g_0$ provided by $[g_{0},g_{1}]\subset
g_{1}$. First of all, if the perturbation is to be given by a single primary
field, then this representation has to be irreducible. This amounts to
restrict the choice of $G/G_0$ to the, so called, `irreducible symmetric
spaces'~\cite{helgas}, which have been completely classified by Cartan and
are labelled by type~I and type~II. 

In Section~\ref{SSSG}, we summarize the main features of the SSSG theories.
In Section~\ref{typeI}, the conformal dimension of the perturbation
corresponding to all the SSSG models related to type~I symmetric spaces is
calculated by making use of the relationship between the classification of
type~I symmetric spaces and the classification of the finite order
automorphisms of complex Lie algebras. It is worth noticing that this
analysis only depends on the structure of the representation of $g_0$ given
by $[g_{0},g_{1}]\subset g_{1}$. Therefore, our results apply to any SSSG
related to a type~I symmetric space irrespectively of the choice of the
normal subgroup $H$ that determines the coset $G_0/H$ and specifies the
underlying CFT. For example, they provide the conformal dimension of the
perturbation in the SSSG models constructed by Bakas, Park and Shin
in~\cite{sssg3}, which include the generalized sine-Gordon models related
to the NAAT equations based on $sl(2)$ embeddings constructed by Hollowood,
Miramontes and Park in~\cite{ls}.

The classical integrability of all these theories is a consequence of the
relationship between their equations of motion and the NAAT equations, which
ensure the existence of an infinite number of classically conserved quantities.
Concerning quantum integrability, it can be established by invoking a
well known result due to Parke, which affirms that the existence of two higher
spin conserved quantities of different spin in a two-dimensional quantum field
theory is enough to ensure that there is no particle production in scattering
processes and that the $S$-matrix is factorizable~\cite{parke}. This way,
the quantum integrability of the HSG theories was demonstrated in~\cite{hsg1} by
checking that the classically conserved quantities of spin $\pm2$ remain
conserved in the quantum theory after an appropriate renormalization. The
majority of SSSG theories also have classically conserved quantities of spin
$\pm2$, and we expect that the analysis in~\cite{hsg1} can be generalized to
those cases without much effort. However, there is a class of SSSG where the
simplest higher spin classically conserved quantities are of spin $\pm3$: the
theories related to symmetric spaces of maximal rank, {\it i.e.\/}, those which
satisfy ${\rm rank\/}(G/G_0) = {\rm rank\/}(G)$, where the rank of the symmetric
space is the dimension of the maximal abelian subspaces contained in $g_1$. In
Section~\ref{integra}, we explicitly check that at least two of
those classically conserved quantities of spin $\pm3$ remain conserved 
in the quantum theory for all the theories related to
a symmetric space of maximal rank where~$H$ is
trivial. We name these theories `Split models', and they are the only SSSG
theories which correspond to massive quantum integrable perturbations of WZNW
theories. It is worth noticing that (marginal) perturbations of WZNW models
have been recently considered in the context of ${\rm AdS}_3$ black
holes~\cite{Emparan}. 

The results of Section~\ref{integra}, together with those of~\cite{hsg1},
support the conjecture that all the SSSG are quantum integrable. This
implies that they should admit a factorizable $S$-matrix and the next stage
of analysis consists in establishing its form. Since we expect that it
should be possible to infer the form of the exact $S$-matrix through the
semiclassical quantization of the solitons, in Section~\ref{soliton} we
investigate the general features of the spectrum of solitons. In view of
the different kinds of SSSG theories, we have restricted the analysis to
the Split models, which illustrate the main properties to be expected. Like
the sine-Gordon and complex sine-Gordon theories, the fundamental particles
of the theory can be identified with some of the classical soliton
solutions. Moreover, the solitons of the Split models are topological and
do not carry Noether charges, again like the solitons of the sine-Gordon
theory. The origin of the topological charge is both the existence of
different vacua and the fact that $G_0$ can be non simply connected. This
is in contrast with the solitons of the HSG theories which are not
topological but carry a
$U(1)^{r_g}$ abelian Noether charge~\cite{hsg2}. In general, the solitons
of a generic SSSG theory corresponding to a perturbation of a coset CFT of
the form $G/H$ are expected to carry both topological charges and abelian
Noether charges related to a global symmetry of the classical action
specified by $H$. In this sense, they are analogous to the dyons in
four-dimensional non-abelian gauge theories~\cite{dyon}. Another relevant
feature of the solitons of the Split models, which is expected to be shared
with other SSSG theories, is that their mass spectrum suggests that some of
them might describe unstable particles in the quantum theory, which is
analogous to what happens in the HSG theories~\cite{hsg2,hsg3}. Actually,
the unstability of the heaviest solitons in the spectrum has been checked by
Brazhnikov in the SSSG theories related to $SU(3)/SO(3)$.

Our conclusions are presented in Section~\ref{conclusions}, and we
have collected the explicit expressions for the classically conserved densities
of the Split models together with some useful algebraic notation in the
Appendix.

\sect{The Symmetric Space sine-Gordon theories.}
\ \indent\label{SSSG}

The different SSSG models of~\cite{ntft} are characterized by
four algebraic data: $\{g,\sigma,\Lambda_\pm\}$. $g$ is a compact
semisimple finite Lie algebra, and $\sigma$ is an involutive ($\sigma^2=1$) 
automorphism of $g$ that induces the decomposition
\be
g=g_0 \oplus g_1,
\lab{decomp}
\ee
with $g_0$ the set of fixed points of $\sigma$ and $g_1=\{u\in g \mid
\sigma(u)=-u\}$. The subspaces $g_0$ and $g_1$ satisfy the
commutation relations~\rf{gradation}, which exhibits that $G/{G_0}$ is a compact
symmetric space, where $G$ and $G_0$  are the  Lie groups corresponding to the
compact Lie algebras $g$ and $g_0$, respectively. Actually, the different
choices for $\{g,\sigma\}$ are in one-to-one relation with the different compact
symmetric spaces $G/G_0$, which prompted the name chosen in~\cite{ntft} for this
class of models.

Finally, $\Lambda_+$ and $\Lambda_-$, are two semisimple
elements in $g_1$ whose choice is only restricted by the condition that
\be
{\rm Ker\/}({\rm ad\/}_{\Lambda_+}) \cap g_0 = {\rm Ker\/}({\rm
ad\/}_{\Lambda_-}) \cap g_0 = g_0^0
\lab{lambcond}
\ee
is abelian, which is required to ensure the existence of a mass gap~\cite{ntft}.
They play the role of continuous coupling constants. 

The SSSG model associated to $\{g,\sigma,\Lambda_\pm\}$ or, equivalently, to
$\Lambda_\pm$ and the symmetric space $G/G_0$ is specified by the action
\be
S_{\rm SSSG}\,=\,\frac{1}{\b^2}\,\left\{\,
{S_{\rm WZNW}[h]}\,+\frac{m^2}{\pi}\,\int{\,d^2 x
\,\langle\,{\Lambda }_ {+},h^{\dagger}{\Lambda} _{-}h\,\rangle}\right\},
\lab{action}
\ee
where $h$ is a bosonic field taking values in $G_0$, ${S_{\rm WZNW}}$ is the
gauged WZNW action corresponding to the coset ${G_0}/H$,
with $H$ the abelian Lie group corresponding to $g_0^0$, and
$m$ is the only mass scale of the theory~\cite{ntft}. The SSSG
action~\rf{action} is invariant with respect to the abelian gauge
transformations
\be
h\rightarrow e^{\a} h e^{-\tau(\a)} 
\lab{field}
\ee
for any $\alpha=\alpha(x,t)\in g_0^0$. Since the potential $V(h)=-m^2\>
\langle\,{\Lambda }_ {+},h^{\dagger}{\Lambda} _{-}h\,\rangle/\pi$ introduced in 
\rf{action} has $H\times H$ left-right symmetry, this gauge
symmetry is essential to make the SSSG theories exhibit a mass gap. 
The precise form of the group of gauge transformations is specified by $\tau$,
which is an automorphism of $g_0^0$ that will not play any
role in the following sections; we refer the reader to~\cite{ntft} for 
further details about the conditions to be satisfied by $\tau$. 
The gauge symmetry leaves a residual global $H=U(1)^p$-symmetry, with $p= {\rm
dim\/}(g_0^0)$, which, in particular, makes the classical solitonic solutions
carry conserved abelian Noether charges, a feature that is shared with the HSG
theories. Moreover, as will be discussed in Section~\ref{soliton}, the
solitonic solutions of the SSSG theories may also carry topological charges.
This possibility does not exist in the case of the HSG theories where all the
possible vacuum configurations get identified modulo gauge transformations. 

The action~\rf{action} can be obtained by Hamiltonian reduction of a
gauged two loop WZNW model~\cite{Luiz,twoloop} associated to the affine
Kac-Moody algebra
$\bar{g}^{(r)}$, where $\bar{g}$ is the complexification of $g$ and
$r=1,2,$ or~$3$ is the least positive integer for which $\sigma^r$ is an
inner automorphism (see the comments below~\rf{loop}).
In fact, the Hamiltonian reduction approach has been recently followed by
Gomes {\it et al.\/} to construct another class of (classical) affine
non-abelian Toda models different to the SSSG theories~\cite{Gomes}. 

The classical integrability of these theories is a consequence of the
connection between their equations of motion and the non-abelian affine Toda
equations~\cite{nt,ls}. This is made explicit by considering a particular gauge
fixing prescription where the classical equations of motion reduce
to~\cite{ntft} 
\bea
&&{\partial}_{-}{(h^{\dag}{\partial}_{+}h)}\,=\,-m^2\,
[\,\Lambda_{+}\,,\,h^{\dag}{\Lambda_{-}}h\,], \\
&&P(h^{\dag}\partial_{+}h)\,=
\,P(\partial_{-}hh^{\dag})\,=\,0,
\lab{equation12} 
\ena
where $P$ is a projector onto the subalgebra $g_0^0$, and $x_\pm = t\pm x$ are
the light-cone variables. The first equation in~\rf{equation12} is a non-abelian
affine Toda equation, and the second provide a set of constraints which 
come from the variation of the action \rf{action} with  respect to the abelian
gauge connections in the LS gauge~\cite{ntft}. Notice that the
equations~\rf{equation12} are left invariant by the transformation
\be
x \ra -x\>, \quad h\ra h^\dagger\>, \quad \Lambda_\pm \ra \eta^{\pm1}
\Lambda_\mp\>,
\lab{parity}
\ee
for an arbitrary real number $\eta$, which shows that the SSSG theories are
parity invariant only if $\Lambda_+ = \eta \Lambda_-$ for some real number
$\eta$, {\it i.e.\/}, if $\Lambda_+$ and $\Lambda_-$ are chosen to be parallel
or anti-parallel~\cite{ntft,ls}.

At the quantum level, the SSSG theories will be described as perturbed CFT's of
the form
\be
S=S_{\rm CFT} + {m^2\over \pi \beta^2} \int{d^2} x \> \Phi (x,t).
\lab{pertur}
\ee
If, according to~\rf{lambcond}, $g_0^0= u(1)^p$ with $p\geq0$, $S_{\rm
CFT}$ will be the action of either the CFT associated to the coset $G_0/U(1)^p$
($p\not=0$) or the WZNW model corresponding to $G_0$ ($p=0$). Moreover, the
perturbation is given by $\Phi =\langle\,{\Lambda }_ {+},h^{\dagger}{\Lambda}
_{-}h\,\rangle$, which will be understood as a matrix element of the WZNW field
taken in the representation of $G_0$ provided by
$[g_0, g_1]\subset g_1$. As shown originally by Bakas for the complex 
sine-Gordon theory~\cite{CSGBAK}, and used in~\cite{hsg1} to define the HSG
theories at the quantum level, these identifications constitute a 
non-perturbative definition of the SSSG theories.

For a given symmetric space $G/G_0$, it is important to emphasise that the form
of the coset ${G_0}/U(1)^p$ is fixed by the choice of $\Lambda_\pm$. Since
$\Lambda_\pm$ are semisimple elements of $g$, different choices will lead to
different values of $p$ in the range
\be
0 \leq {\rm rank\/}(G)-{\rm rank\/}(G/G_0)  \leq p \leq {\rm min\/}\Bigl[{\rm
rank\/}(G_0)\>, \> {\rm rank\/}(G)- \nu\Bigr],
\lab{dim}
\ee
where the rank of the symmetric space, ${\rm rank\/}(G/G_0)$, is the
dimension of the maximal abelian subspaces contained in $g_1$, and $\nu=2$
or~1 depending on whether $\Lambda_+$ and $\Lambda_-$ are linearly independent or
not, respectively. In particular, the lower bound is reached when
$\Lambda_+, \Lambda_- \in g_1$ are regular and, hence, ${\rm Ker\/}({\rm
ad\/}_{\Lambda_\pm})$ is already a maximal abelian subspace of $g$. All this
implies that the SSSG theories provide a rich variety of different
integrable models that include, for $p = {\rm rank\/}({G_0})$, new  massive
perturbations of the theory of $G_0$-parafermions different than those
provided by the Homogeneous  sine-Gordon theories~\cite{hsg1}. Notice that
this case happens only if the symmetric space satisfies ${\rm
rank\/}(G_0)\leq {\rm rank\/}(G)- \nu$. Another particularly interesting
class of models occurs when ${\rm rank\/}(G)={\rm rank\/}(G/{G_0})$ and
$p=0$. In this case, the SSSG theory is just a massive perturbation of the
WZNW model corresponding to $G_0$.

We will concentrate on the theories where the perturbation in~\rf{pertur} is 
given by a single spinless primary field of the~CFT which, taking into
account the properties of the WZNW field~\cite{wzw}, amounts to restrict
ourselves to the SSSG theories associated with symmetric spaces where the
representation of $g_0$ provided by $[g_0,g_1]\subset g_1$ is irreducible.
Otherwise,  the perturbation will be the sum of more than one primary
field. The symmetric spaces with that property are called `irreducible',
and have been completely classified by Cartan~\cite{helgas}. There are two
types of compact irreducible symmetric  spaces:
\begin{itemize}
\item
{\em Type I,} where the compact Lie algebra $g$ is simple.
\item
{\em Type II,} where the compact Lie algebra is of the
form $g=g_1 \oplus g_2$ with $g_1=g_2$ simple, and the involution $\sigma$
interchanges $g_1$ and $g_2$.
\end{itemize}
In the following we will only consider the SSSG theories associated to the 
type~I symmetric spaces, which admit a thorough classification. 

\sect{The type~I SSSG theories as perturbed CFT's.}
\ \indent\label{typeI}

In this section, we calculate the central charge of the unperturbed CFT and
the conformal dimension of the perturbation corresponding to
the SSSG theories associated with the symmetric spaces of type~I. We will make 
use of the relationship between the Cartan classification of this type of
symmetric spaces and the Kac classification of the automorphisms of finite
order of complex Lie algebras, which provides a systematic and very convenient
description of the involutive automorphism~$\sigma$. This represents an
important advantage with respect to previous works on integrable systems
associated with symmetric spaces~\cite{sssg3,sssg1} which
generally make use of explicit parametrizations of the field~$h$ based on
some matrix representation for the symmetric space. 

\subsection{Type~I symmetric spaces and finite order automorphisms.}

\ \indent\label{typeI1}
The basic result is due to Kac, who established the following correspondence
between the involutions of a complex Lie algebra $\bar{g}$ and 
the involutions of its compact real form $g$. 

\begin{theo} 
\label{theo1}
{\rm (Proposition 1.4 in~\cite{helgas}, Ch.~X)} Let $Aut(g)$ denote
the set of automorphisms of $g$, $Inv(g)$
the subset containing the involutions, and ${Inv(g)}/{Aut(g)}$ the set of
conjugacy classes in $Aut(g)$ of the elements in $Inv(g)$. We define 
${Inv(\bar{g})}/{Aut(\bar{g})}$ similarly. Each automorphism
$\sigma\,\in \, Inv(g)$ extends uniquely to $\bar\sigma\,\in \, 
Inv(\bar{g})$ and if $\sigma_1$, $\sigma_2$ are conjugate within
$Aut(g)$, then $\bar\sigma_1$, $\bar\sigma_2$ are conjugate
within $Aut(\bar{g})$. Taking into account all this, it can be proved that the
mapping:
\be
\tau:Inv(g)/Aut(g) \,\rightarrow\,
Inv{(\bar{g})}/Aut{(\bar{g})},
\lab{clasi8}
\ee
induced by $\sigma\,\rightarrow \,
\bar{\sigma}$ is a bijection.
\end{theo}

Recall now that the different compact symmetric spaces of type~I associated with 
a compact simple Lie algebra~$g$ are in one-to-one relation with the different
involutive automorphisms of~$g$, not distinguishing automorphisms which are
conjugate by the group $Aut(g)$. Therefore, they are also in one-to-one
relation with the involutive automorphisms of its complexification $\bar{g}$,
modulo conjugations by $Aut{(\bar{g})}$. 

In order to summarize the Kac classification of the
automorphisms of finite order of complex Lie algebras, it is necessary to 
introduce the following notation (see~\cite{kac}, Ch.~8, for more details). Let
$\bar{g}$ denote a complex simple Lie algebra and $\mu$  an automorphism of
$\bar{g}$ induced by an automorphism of its Dynkin diagram of order $r=1,2$ or
$3$. $\mu$  induces a ${\ZZ}/{r\ZZ}$--gradation of $\bar{g}$, which means that
$\bar{g}$ can be decomposed into the sum of a set of subspaces
labelled by an integer $0 \leq k \leq r-1$ that satisfy
\bea
&&\mu(u)\> =\> {\rm e\>}^{{2\pi i\over r}\>j}\> u \quad \forall u\in
\bar{g}_j(\mu)\>,
\nn
&&\bar{g}=\bigoplus_{k=1}^{r-1} \bar{g}_k(\mu), \qquad [\bar{g}_j(\mu),
\bar{g}_k(\mu)]\subset \bar{g}_{j+k\; {\rm mod}\; r\>}(\mu).
\ena
Then there is a particular set of generators of
$\bar{g}$, $\{E_0,E_1, \ldots,E_l\}$, where $l$ is the rank of the invariant
subalgebra $\bar{g}_0{(\mu)}$, with the following properties:

\begin{itemize}
\item[a)] $\{E_1,\ldots,E_l\}\in \bar{g}_0{(\mu)}$ for
$(\bar{g},r)\,\neq \,(A_{2l},2)$, and
$\{E_0, E_1,\ldots,E_{l-1}\}\in \bar{g}_0{(\mu)}$ for
$(\bar{g},r)\,= \,(A_{2l},2)$.

\item[b)] If $(\bar{g},r)\,\neq \,(A_{2l},2)$, 
then $E_0\,\in \,\bar{g}_1{(\mu)}$ is the lowest--weight vector of the 
irreducible representation of $\bar{g}_0{(\mu)}$ given by $
[\,\bar{g}_0{(\mu)},\bar{g}_1{(\mu)}\,]
 \,\subset\,\bar{g}_1{(\mu)}$. 
Conversely, when $(\bar{g},r)\,= \,(A_{2l},2)$ this role is played by $E_l$.

\item[c)] $\{E_1,\ldots,E_l\}$ are positive Chevalley generators for
$\bar{g}_0{(\mu)}$, except for
$(\bar{g},r)=(A_{2l},2)$ where the Chevalley generators are
$\{E_0, E_1,\ldots,E_{l-1}\}$.

\item[d)] Results a), b), and c) correspond
to the case $r>1$. When $r=1$, $l={\rm rank\/}(\bar{g})$, $E_1\,=\,E_{\a_1}, 
\ldots, E_l\,=\,E_{\a_l}$, and $E_0=E_{-\Psi_g}$, where 
$\{\vec{\a}_1, \vec{\a}_2, \ldots, \vec{\a}_l\}$ is a set of simple roots of
$\bar{g}$, and $\vec{\Psi}_g$ is the highest root.
\end{itemize}
The classification is given by the following theorem.

\begin{theo} 
\label{theo2}
{\rm (Theorem 8.6 in~\cite{kac})} 
Let $\vec{s}=(s_0, s_1,\ldots, s_l)$ be a sequence of 
non-negative relatively prime integers, and put 
\be
m = r\sum_{i=0}^l a_i s_i
\lab{order}
\ee
where $a_0,a_1,\ldots, a_l$ are the Kac labels corresponding to the Dynkin
diagram of the (twisted if $r\not=1$) affine Kac-Moody algebra $\bar{g}^{(r)}$.
Then:

\begin{itemize}
\item[a)] The relations 
\be
\sigma_{\vec{s};r}(E_j)=e^{{2 \pi i {s_j}}/m}  E_j, \quad j=0,\ldots, l,
\ee
define uniquely an $m$-th order automorphism $(\vec{s};r)$ of $\bar{g}$.

\item[b)] Up to conjugacy by an automorphism of $\bar{g}$,
the automorphisms $\sigma_{\vec{s};r}$ exhaust all $m$-th
order automorphisms of  $\bar{g}$.

\item[c)] The elements $\sigma_{\vec{s};r}$ and
$\sigma_{\vec{s}';r'}$  are conjugate by an automorphism
of $\bar{g}$ if, and only if, $r=r'$ and the sequence $\vec{s}$
can be transformed into the sequence  $\vec{s}'$
by an automorphism of the Dynkin diagram of $\bar{g}^{(r)}$.
\end{itemize}\end{theo}

Taking into account Theorems~\ref{theo1} and~\ref{theo2}, the classification of
the symmetric spaces of type~I is equivalent  to working out the equation
\be
m\,=\,r\sum_{i=0}^{l}{a_i}{s_i}\,=\,2,
\lab{grad}
\ee
which has only three possible types of solutions:
\bea
&&[{\rm A1}]\quad r=1\>, \>a_{i_0}=2\>, \>s_{i_0}=1\>, \quad {\rm and}\quad 
s_i=0
\quad{\rm for}\quad i \neq i_0\>, \nn
&&[{\rm A2}]\quad r=2\>, \>a_{i_0}=1\>, \>s_{i_0}=1\>, \quad{\rm and}\quad
s_i=0 \quad{\rm for}\quad i\neq i_0\>, \nn
&&[B]\>\quad r=1\>, \>a_{i_0}=a_{i_1}=1\>, \>s_{i_0}=s_{i_1}=1\>,
\quad {\rm and}
\quad s_i=0\quad {\rm for}\quad i\neq i_0\>, \> i_1.
\lab{AAB}
\ena
This classifies the type~I (compact) symmetric spaces and, hence, the
corresponding SSSG theories, into type A1, A2, and B~\cite{helgas}.

Given an automorphism $\sigma_{\vec{s};r}$, the subspaces
$\bar{g}_0$ and $\bar{g}_1$ can be easily characterized as follows.

\begin{theo} 
\label{theo3}
{\rm (Proposition 8.6 in~\cite{kac})}
\begin{itemize}
\item[a)] Let $i_1,i_2,\ldots,i_p$ be all the indices
for which $s_{i_1}=\cdots=s_{i_p}=0$.
Then the Lie algebra  $\bar{g}_0$ is isomorphic to a direct
sum of the ({\it l-p})--dimensional centre and a semisimple
Lie algebra whose Dynkin diagram is the subdiagram of the
Dynkin diagram of $\bar{g}^{(r)}$ consisting of
the vertices $i_1,\ldots,i_p$.

\item[b)] Let $j_1,\ldots,j_n$ be all the indices for
which  $s_{j_1}=\cdots=s_{j_n}=1$. Then the 
representation of  $\bar{g}_0$ provided by $[\bar{g}_0,\bar{g}_1] \subset
\bar{g}_1$ is isomorphic to
a direct sum of $n$ irreducible modules with
highest weights $-\vec{\a}_{j_1},\ldots,-\vec{\a}_{j_n}$.
\end{itemize}
\end{theo}
Eq.~\rf{AAB} and Theorem~\ref{theo3} imply that $g_0$ is of the form
\be
g_0\> =\> \cases{\bigoplus_{i=1}^q g^{(i)}& for type A1 and A2,\cr
\noalign{\vskip 0.2truecm}
\bigoplus_{i=1}^q g^{(i)} \oplus u(1)& for type B,}
\lab{ABform}
\ee
where $q$ can be either~1 or~2 in both cases, and $g^{(i)}$ is
always compact and simple. 

Therefore, if the symmetric space is of type A1 or A2, $S_{\rm CFT}$ 
in~\rf{pertur} is the action of the CFT associated to a coset of the form
\be
\bigotimes_{i=1}^q {G^{(i)}_{k_i}\> /\> U(1)^p}\>,
\lab{coset}
\ee
where $G^{(i)}$ is the compact simple Lie group corresponding to $g^{(i)}$, and
$p$ is some integer in the range~\rf{dim}. Since $g_0$ is non-abelian, the
consistency of the quantum theory requires that the coupling constant
in~\rf{action} is quantized. The precise form of the quantization rule
is~\cite{wzw,olive}
\be
{1\over \hbar \beta^2} \> =\> {\vec{\Psi}_{g^{(i)}}^2\over 2}\> k_i\>,
\lab{level}
\ee
where $k_i$ is an integer for each simple factor in~\rf{ABform},
and $\vec{\Psi}_{g^{(i)}}^2$ is the square length of the long
roots of $g^{(i)}$ with respect to the bilinear form of $g$.
In~\rf{level} we have shown explicitly the Plank constant to exhibit
that, just as in the sine-Gordon theory, the semi-classical limit is the
same as the weak coupling limit, and that both are recovered when
$k_i\rightarrow \infty$. Since there is a unique coupling constant
$\beta^2$, eq.~\rf{level} implies that the `levels' in~\rf{coset}
are related by means of
\be
{k_i\over  k_j} \> =\> {\vec{\Psi}_{g^{(j)}}^2 \over \vec{\Psi}_{g^{(i)}}^2}\>.
\lab{levelPlus}
\ee
Nevertheless, our calculations show that
$\vec{\Psi}_{g^{(1)}}^2=\vec{\Psi}_{g^{(2)}}^2$ for all the type~I symmetric
spaces, with the only exception of $G_2/SU(2)\times SU(2)$ where
$\vec{\Psi}_{g^{(2)}}^2=3
\vec{\Psi}_{g^{(1)}}^2$ (see Section~\ref{G2}), and we will use the notation
$k=k_1$ in the following.

Therefore, in this case, the central charge of the unperturbed CFT is
\be
c_{\rm CFT}=\sum_{i=1}^{q} \frac{k_i \> {\rm dim}(g^{(i)})}
{k_i+ h_i^{\lor}}\> -\> p,
\lab{centralA}
\ee
where $h_{i}^{\lor}$ is the dual Coxeter number of $g^{(i)}$.
Since the perturbation $\Phi$ is just a matrix
element of the WZNW field in the representation of $G_0$ provided by
$[g_0, g_1]\subset  g_1$, which is irreducible, $\Phi$ is a spinless primary
field with conformal dimension~\cite{wzw,olive}
\be
\Delta_\Phi = \overline{\Delta}_\Phi = \sum_{i=1}^q {C_2(g^{(i)})/
\vec{\Psi}_{g^{(i)}}^2 \over k_i+ h_{i}^{\lor}},
\lab{dimensionfiA}
\ee
where $C_2(g^{(i)})$ is the quadratic Casimir. Then, Theorem~\ref{theo3} shows
that this representation is a highest weight representation and, hence, the
quadratic Casimir is given by
\be
C_2(g^{(i)})\> =\> \langle\,\vec{\Lambda}\> ,\>  \vec{\Lambda}\,+
\,2\vec{\d}^{(i)}\,\rangle
\lab{Casimir}
\ee
where $\vec{\Lambda}= -\vec{\alpha}_{i_0}$ is the highest weight, and
$\vec{\d}^{(i)}$ is half the sum of the positive roots of $\bar{g}^{(i)}$.

Let us consider now the SSSG theories associated with symmetric spaces of
type~B. The main difference with the SSSG models of type~A1 or~A2 is that $g_0$
includes now a one-dimensional centre: the $u(1)$ factor in~\rf{ABform}. In
this case, it will be convenient to choose the Cartan subalgebra of
$g$ such that it contains the Cartan subalgebras of $\bigoplus_{i=1}^q
g^{(i)}$ in addition to the generator of the centre. Then, $u(1) = \RR\>
i\vec{u}\cdot
\vec{h}$, where $\vec{u}$ is a vector that is orthogonal to all  the roots
of $\bigoplus_{i=1}^q g^{(i)}$, and the components of $i\vec{h}$ provide a
basis for the Cartan subalgebra of $g$.  According to~\ref{theo3}, another
important difference is that the representation of $\bar{g}_0$ given by
$[\bar{g}_0, \bar{g}_1] \subset
\bar{g}_1$ is the sum  of two irreducible highest weight representations
with  highest weights $\vec{\Lambda}_1= -\vec{\alpha}_{i_0}$ and
$\vec{\Lambda}_2= -\vec{\alpha}_{i_1}$, namely, 
\be
\bar{g}_1 \>=\> L(\vec{\Lambda}_1) \>\oplus \> L(\vec{\Lambda}_2)\>.
\ee
Let us consider the identity
\be 
\sum_{i=0}^l a_i \> \vec{\alpha}_i \> =\> 0
\lab{Kaclabel}
\ee
satisfied by the Kac labels of the Dynkin diagram of $\bar{g}^{(1)}$, where
$\{\vec{\a}_1, \vec{\a}_2, \ldots, \vec{\a}_l\}$ is a set of simple roots of
$\bar{g}$, and $\vec{\Psi}_g$ is the highest root. Then, the two
highest weights satisfy
\be
\vec{\Lambda}_1\> +\> \vec{\Lambda}_2\> =\> \sum_{i=0\atop i\not= i_0, i_1}^l 
a_i\>
\vec{\alpha}_i \>,
\lab{conjugate}
\ee
which implies that $\vec{\Lambda}_1$ is minus the lowest weight of the
representation with highest weight $\vec{\Lambda}_2$. This manifests that the two
highest weight representations are conjugate, $L(\Lambda_2) = L^\dagger 
(\Lambda_1)$, which is consistent with the fact that the representation
of $g_0$ given by $[g_0, g_1]\subset g_1$ is irreducible. Therefore, since
$\Lambda_\pm \in g_1$, it can be decomposed as
\be
\Lambda_\pm \> =\> \lambda_\pm \> +\> \lambda_\pm^\dagger\>, \quad {\rm with} 
\quad \lambda_\pm \in L(\Lambda_1)\>.
\lab{LambComp}
\ee
Taking again into account Theorem~3, eq.~\rf{conjugate} also implies that
$\vec{\Lambda}_1+ \vec{\Lambda}_2$ is a linear combination of the roots of
$g_0$, which are orthogonal to $\vec{u}$, and, hence,
\be
\vec{u}\cdot \vec{\Lambda}_1 \> =\> -\> \vec{u}\cdot \vec{\Lambda}_2\>.
\lab{eigenvalue}
\ee
Our calculations show that, in all the symmetric spaces of type~B,
$\vec{u}\cdot \vec{\Lambda}_1\not=0$. Since $\Lambda_\pm \in g_1$, this 
implies that $[\vec{u}\cdot \vec{h}, \Lambda_{\pm}]\not=0$ and, hence, the
$u(1)$ is not in $g_0^0$ for any choice of $\Lambda_\pm$ (see~\rf{lambcond}). 

Consider a generic field configuration
\be
h\> =\> \widetilde{h}\> \exp \bigl(i \varphi \, \vec{u}\cdot\vec{h}\bigr)\>,
\ee
where $\widetilde{h}$ is a field taking values in the compact semisimple Lie 
group $\bigoplus_{i=1}^q G^{(i)}$, and $\varphi=\varphi(x,t)$ is
a real scalar field; for convenience we will normalize $\vec{u}$ such that
$\vec{u}\cdot\vec{u}= 4\pi$. Then, the action~\rf{action} becomes
\bea
&& S_{\rm SSSG}= \frac{1}{\b^2}\,\biggl\{
S_{\rm WZNW}[\widetilde{h}]\> +\> {1\over2}\> \int d^2x\>
\partial_\mu \varphi\> 
\partial^\mu \varphi \nn
\noalign{\vskip 0.2truecm}
&&\qquad +\frac{m^2}{\pi}\,\int d^2 x \left(
{\rm e\>}^{-i \varphi \, \vec{u}\cdot\vec{\Lambda}_1} \>\langle\,{\lambda
}_{+}^\dagger,\widetilde{h}^{\dagger}{\lambda}_{-}\widetilde{h}\,\rangle\> + \>
{\rm e\>}^{+i \varphi \, \vec{u}\cdot\vec{\Lambda}_1} \>\langle\,{\lambda
}_{+},\widetilde{h}^{\dagger}{\lambda}_{-}^\dagger\widetilde{h}\,\rangle\right)
\biggr\}\>.
\lab{actionB}
\ena
Therefore, if the symmetric space is of type~B,
$S_{\rm CFT}$ in~\rf{pertur} is the action of the CFT associated to a coset of
the form
\be
\Bigl[\bigotimes_{i=1}^q {G^{(i)}_{k_i}\> /\> U(1)^p}\Bigr] \times U(1)\>,
\lab{cosetB}
\ee
{\it i.e.\/}, a coset of the form~\rf{coset} plus a massless boson, whose
central charge is
\be
c_{\rm CFT}=\sum_{i=1}^{q} \frac{k_i \> {\rm dim}(g^{(i)})}
{k_i+ h_i^{\lor}}\> +\> 1\> -\> p,
\lab{centralB}
\ee
with the levels $k_i$ defined by the quantization rule~\rf{level}.
In~\rf{cosetB}, we have already taken into account that the centre of $g_0$ is
not in $g_{0}^0$ as a consequence of~\rf{LambComp} and~\rf{eigenvalue}.
Therefore, in this case, ${\rm rank\/}(G_0)$ has to be substituted for ${\rm
rank\/}(G_0)-1$ on the right-hand-side of eq.~\rf{dim}. Concerning the
perturbation
$\Phi$ in~\rf{actionB}, it is a primary field of conformal dimension
\be
\Delta_\Phi = \overline{\Delta}_\Phi = \sum_{i=1}^q {C_2(g^{(i)})/
\vec{\Psi}_{g^{(i)}}^2 \over k_i+ h_{i}^{\lor}}
\> +\> {(\vec{u}\cdot\vec{\Lambda}_1)^2\over k\> \vec{\Psi}^2_{g^{(1)}}} 
\>,
\lab{dimensionfiB}
\ee
where the quadratic Casimir is given by~\rf{Casimir} with $\vec{\Lambda}=
\vec{\Lambda}_1$ or $\vec{\Lambda}_2$, both leading to identical results.

At this stage, we would like to correct a wrong statement in~\cite{ntft}
concerning the fields associated to the centre of $g_0$. In that article, it
was said that those fields can always be decoupled whilst preserving
integrability. However, the SSSG of type~B shows that this is not true in
general. In our case, there is only one field associated to the centre of
$g_0$, $\varphi=\varphi(x,t)$, whose classical equation of motion is
\be
\partial_+\partial_-\varphi \> =\> -\> i\> m^2\> {\vec{u}\cdot \vec{\Lambda_1}
\over 4\pi}\> 
\left({\rm e\>}^{-i \varphi \, \vec{u}\cdot\vec{\Lambda}_1} \>\langle\,{\lambda
}_{+}^\dagger,\widetilde{h}^{\dagger}{\lambda}_{-}\widetilde{h}\,\rangle\> - \>
{\rm e\>}^{+i \varphi \, \vec{u}\cdot\vec{\Lambda}_1} \>\langle\,{\lambda
}_{+},\widetilde{h}^{\dagger}{\lambda}_{-}^\dagger \widetilde{h}\,\rangle
\right)\>,
\ee
which clearly shows that $\varphi$ cannot be decoupled simply by putting
$\varphi(x,t)=0$ unless
$\langle\lambda_+^\dagger,\widetilde{h}^{\dagger}\lambda_-\widetilde{h}
\rangle$ is real, which is equivalent to 
\be
[\Lambda_+, \widetilde{h}^{\dagger}\Lambda_-\widetilde{h}] \in
\bigoplus_{i=1}^q g^{(i)}\>.
\ee
This condition was already noticed in~\cite{ls}.

There are two general features of the conformal dimensions given
by~\rf{dimensionfiA} and~\rf{dimensionfiB} that is important to emphasize.
The first one is that the conformal dimension
of the perturbation is independent of the value of $p$ in eqs.~\rf{coset}
and~\rf{cosetB}, which is a consequence of the fact that the potential
in~\rf{action} and, hence, $\Phi$ are invariant with respect to the gauge
transformations~\rf{field}. The second is that $\Delta_\Phi$ decreases with
$k$, which means that the perturbation is always relevant for $k$ above some
minimal value characteristic of each SSSG theory. Actually, $\Delta_\Phi$
vanishes when $k\rightarrow \infty$, which shows that
the theory consists of ${\rm dim\/}(g_0) - p$ bosonic massive particles
in the the semi--classical and/or weak coupling limit.

\subsection{Explicit calculation of $\Delta_\Phi$.}

\indent \ \

In the following we will illustrate the general procedure to calculate the
conformal dimension of the perturbation by
considering three particular cases where $g_0$ is either simple
($SU(2n)/SO(2n)$), semisimple ($G_2/SU(2)\times SU(2)$), or the direct  sum of
a simple ideal and a one-dimensional centre ($Sp(n)/U(n)$). In all these
examples, ${\rm rank\/}(G/G_0) = {\rm rank\/}(G_0)$, which, according
to~\rf{dim}, means that $\Lambda_\pm$ can be chosen such that $p=0$
in~\rf{coset} and~\rf{cosetB}. The results for all the type~I symmetric 
spaces are presented in table~\ref{table1} and~\ref{table2}. 
Other useful features of the symmetric spaces of type~I have been collected
in tables~\ref{table3}--\ref{table5}. 
In these tables, we have already taken into account the following 
isomorphisms of Lie algebras:
$$
su(2)\simeq so(3) \simeq sp(1)\>, \quad
so(5)\simeq sp(2)\>, \quad so(4)\simeq su(2)\oplus su(2)\>, \quad
su(4)\simeq so(6)\>.
$$
In particular, this shows that $SO(4)$ is not simple and, therefore, no symmetric
space with $G=SO(4)$ appears in the tables because it would not be of type~I.
Moreover, the symmetric space $SU(2)/SO(2)$ corresponds to the well known
sine-Gordon theory where the field takes values in the abelian group $SO(2)\simeq
U(1)$, and it has not been included in the tables.
When $G_0$ is simple, it is worthwhile noticing that
$\Delta_\Phi$ admits the general expression
\be
\Delta_\Phi\> =\> {\vec{\Psi}_g^2\over \vec{\Psi}_{g_0}^2}\> {h_g^\vee\over
2\> (k + h_{g_0}^\vee)}\>.
\ee

\subsubsection{Example I: $G/{G_0}={SU(2n)}/{SO(2n)}$, $n>2$.}

\indent \ \

\begin{figure}[ht]
\begin{center}
\mbox{\epsfig{file=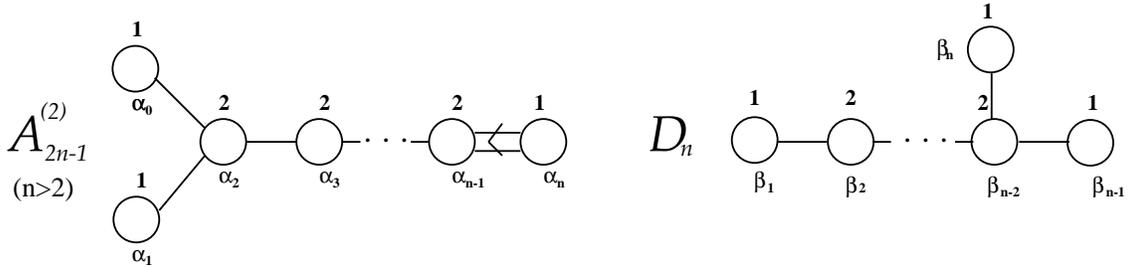, height=3.5cm}}
\caption{Dynkin diagrams of $A_{2n-1}^{(2)}$ and $D_n$. The numbers on
top of the nodes are the Kac labels.}
\label{fig1}
\end{center}
\end{figure}

In this case, $\bar{g}= A_{2n-1}$, $r=2$, and $\vec{s}=(0,\ldots,0,1)$, which
follows from Theorem~\ref{theo3}, part~a), and the observation that the Dynkin
diagram of $\bar{g}_0=D_n$ is a subdiagram of the Dynkin diagram of
$\bar{g}^{(r)}=A_{2n-1}^{(2)}$, as can be seen in fig.~\ref{fig1}. Therefore,
and taking into account~\rf{AAB}, this symmetric space is of type A2.
Fig.~\ref{fig1} also  shows that the roots of $\bar{g}_0=D_n$ are short roots
in the Dynkin diagram of $\bar{g}^{(r)}=A_{2n-1}^{(2)}$, which means that
$\vec{\Psi}_{D_n}^2 = \vec{\Psi}_{A_{2n-1}^{(2)}}^2/2$.

Theorem~\ref{theo3}, part~b), also implies that $[\bar{g}_0,
\bar{g}_1]\subset \bar{g}_1$ is an irreducible representation of
$\bar{g}_0$ with highest weight $\vec{\Lambda}= - \vec{\a}_n$. Then, we can
use the identity~\rf{Kaclabel} in order to write $\vec{\Lambda}$ as a linear
combination of the roots of $\bar{g}_0=D_n$, namely
\be
\vec{\Lambda}\, =\> -\vec{{\a}}_n \,=\, \vec{{\a}}_0\,+\,\vec{{\a}}_1\,+\,
2\sum_{i=2}^{n-1}\vec{{\a}}_i
\,=\> \vec{{\b}}_n\,+\,\vec{{\b}}_{n-1}\,+\,
2\sum_{i=1}^{n-2} \vec{{\b}}_i.
\lab{null3}
\ee
Moreover, taking into account that the highest root of $D_n$ is 
\be
\vec{\Psi}_{D_n}\,=\,\vec{\b}_1\,+\,\vec{\b}_n
\,+\,\vec{{\b}}_{n-1}
\,+\,2\sum_{i=2}^{n-2}{\vec{{\b}}_i},
\lab{longroot}
\ee
eq.~\rf{null3} simplifies to
$\vec{\Lambda}=\vec{\Psi}_{D_n}+\vec{{\b}}_1$. All this allows one to
easily calculate the quadratic Casimir of this highest weight 
representation:
\be
C_2(D_n)\,=\,
\langle \,\vec{\Lambda}\,, \,
\vec{\Lambda}\,+
\,2\vec{\delta}_{D_n}\rangle \,=\,2n \> \vec{\Psi}_{D_n}^2,
\lab{casimir1}
\ee
where we have used the standard realization of
the root system of $D_n$ as a sublattice of the real euclidean
space $\RR^n$:
\be
\Pi_{D_n}\>
=\>\big\{\vec{\beta}_1\,=\,\vec{v}_1-\vec{v}_2\,, \ldots,\vec{\beta}_{n-1}
\,=\,\vec{v}_{n-1}-\vec{v}_{n}  \,,\> \vec{\beta}_n
\,=\,\vec{v}_{n-1}+\vec{v}_{n} \big\}\>,   
\ee
where
\be
\vec{v}_i \cdot \vec{v}_j \>=\> {\vec{\Psi}_{D_n}^2\over 2} \> \delta_{ij}\>,
\ee 
together with 
\be
\vec{\Psi}_{D_n}\, =\, \vec{v}_1 + \vec{v}_2\>, \quad {\rm and} \quad
2\vec{\delta}_{D_n}\> =\> 2\sum_{i=1}^n (n-i)\> \vec{v}_i\>.
\ee

Therefore, taking into account~\rf{centralA} and~\rf{dimensionfiA}, we
conclude that the SSSG's associated with the symmetric space
$SU(2n)/SO(2n)$ are integrable perturbations of either the WZNW 
model corresponding to $SO(2n)$ at level $k$ ($p=0$) or a coset CFT of
the form $SO(2n)_k/U(1)^p$ whose central charge is
\be
c_{\rm CFT\/}= {k\> n\> (2n-1)\over k + 2(n-1)}\> -\> p,
\ee
where $0\leq p\leq n$, and the perturbation has conformal
dimension
\be
\Delta_{\Phi}\,=\,\frac{2n}{k\,+\,2(n\,-\,1)}\>.
\lab{dim1}
\ee
Notice that the perturbation is relevant for $k>2$. 

\subsubsection{Example II: $G/G_0=G_2/SU(2) \times SU(2)$.}
\label{G2}
\indent \ \

\begin{figure}[ht]
\begin{center}
\mbox{\epsfig{file=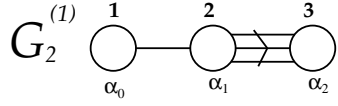, height=1.3cm}}
\caption{Dynkin diagram and Kac labels of $G_2^{(1)}$.}
\label{fig2}
\end{center}
\end{figure}

In this case, $\bar{g}= G_2$, $r=1$, and $\vec{s}=(0,1,0)$, as can be seen in
fig.~\ref{fig2}. Therefore, this symmetric space is of type~A1. 
$\bar{g}_0$ is of the form $\bar{g}_0=\bar{g}^{(1)} \oplus
\bar{g}^{(2)}$ with $\bar{g}^{(1)}=\bar{g}^{(2)}=A_1$. 
$\bar{g}^{(1)}$ and $\bar{g}^{(2)}$ are associated with the roots $\vec{\a}_2$
and $\vec{\a}_0$ of the Dynkin diagram of the affine algebra $G_2^{(1)}$,
respectively, whose length is different and, therefore,
$\vec{\Psi}_{g^{(2)}}^2= 3 \vec{\Psi}_{g^{(1)}}^2 =
\vec{\Psi}_{G_2^{(1)}}^2 $.

Theorem~\ref{theo3} implies that $[\bar{g}_0,
\bar{g}_1]\subset \bar{g}_1$ gives an irreducible representation of
$\bar{g}_0$ with highest weight $\vec{\Lambda}= - \vec{\a}_1$. Then, since
$\vec{\a}_0 + 2\vec{\a}_1 + 3\vec{\a}_2=0$,
one can write 
\be
\vec{\Lambda}\,= \,{3\over2}\>
\vec{\Psi}_{g^{(1)}} \> +\> {1\over2}\> \vec{\Psi}_{g^{(2)}}\>,
\ee
which leads to
\bea
&& C_2(g^{(1)})\,=\,
\langle \,\vec{\Lambda}\,, \,
\vec{\Lambda}\,+ \vec{\Psi}_{g^{(1)}} \,\rangle \,=\,{15\over 4}\>
\vec{\Psi}_{g^{(1)}}^2\>, \nn
\noalign{\vskip0.2truecm}
&& C_2(g^{(2)})\,=\,
\langle \,\vec{\Lambda}\,, \,
\vec{\Lambda}\,+ \vec{\Psi}_{g^{(2)}} \,\rangle \,=\,{3\over 4}\>
\vec{\Psi}_{g^{(2)}}^2\>.
\lab{casimir2}
\ena

Therefore, we
conclude that the SSSG's associated with this symmetric space
are integrable perturbations of either the
WZNW model corresponding to $SU(2)_k\times SU(2)_{3k}$ ($p=0$) or a coset
CFT of the form $SU(2)_k\times SU(2)_{3k}/U(1)^p$ whose central charge is
\be
c_{\rm CFT\/}= \> {3k\over k+2}\>+\> {9k\over 3k+2} -\> p\>.
\ee
Notice the relationship between the levels of the two $SU(2)$ factors, which
is a consequence of~\rf{levelPlus}. The perturbation has
conformal dimension
\be
\Delta_{\Phi}\,=\, {3\over 4(k+2)}\> +\> {15\over 4(3k+2)}\>.
\lab{dim2}
\ee
Moreover, since in this case ${\rm rank\/}(G_0) ={\rm rank\/}(G)=2$,
according to~\rf{dim} there is a different SSSG theory for each integer
$p$ in the range $0\leq p\leq 2-\nu$. The perturbation is relevant for $k>2$.

\subsubsection{Example III: $G/{G_0}\,=\,{Sp(n)}/{U(n)}\,=
\,{Sp(n)}/{U(1) \times SU(n)}$, $n>1$.}

\indent \ \

\begin{figure}[ht]
\begin{center}
\mbox{\epsfig{file=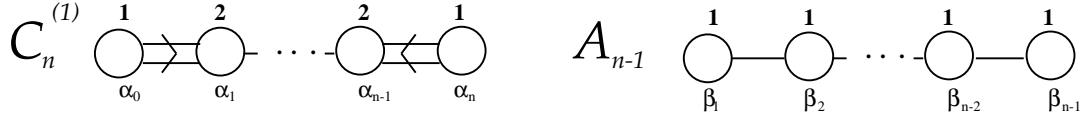, height=1.5cm}}
\caption{Dynkin diagram and Kac labels of $C_n^{(1)}$ and $A_{n-1}$.}
\label{fig3}
\end{center}
\end{figure}

In this case, $\bar{g}= C_n$, $r=1$, and
$\vec{s}=(1,0\ldots,0,1)$, as can be seen in fig.~\ref{fig3}, and the symmetric
space is  of type~B. Then, $\bar{g}_0= \bar{g}^{(1)} \oplus u(1)$ with 
$\bar{g}^{(1)}=A_{n-1}$, and
$\vec{\Psi}_{A_{n-1}}^2 = \vec{\Psi}_{C_n^{(1)}}^2 /2$.

Recall the standard realization of
the root systems of $C_n$ and $A_{n-1}$ as sublattices of the real
euclidean space $\RR^n$:
\bea
&&\Pi_{C_n}\>
=\>\big\{\vec{\a}_1\,=\,\vec{v}_1\,-\,\vec{v}_2\,, \ldots,\vec{\a}_{n-1}
\,=\,\vec{v}_{n-1}\,-\,\vec{v}_{n}  \,,\> \vec{\a}_n
\,=\,2\vec{v}_n \big\}\>, \quad \vec{\Psi}_{C_n}\, =\, -\vec{\a}_0 \,=\,
2\vec{v}_1\>, \nn
&&\Pi_{A_{n-1}}=\big\{ \vec{\a}_1=\vec{v}_1-\vec{v}_2,
\ldots,\vec{\a}_{n-1} =\vec{v}_{n-1}-\vec{v}_{n} \big\} \subset
\Pi_{C_n}\>, \quad \vec{\Psi}_{A_{n-1}}= \vec{v}_1-\vec{v}_{n}\>,
\lab{anroots}
\ena
where
\be
\vec{v}_i \cdot \vec{v}_j \>=\> {\vec{\Psi}_{C_n}^2\over 4} \> \delta_{ij}\>.
\ee 

Using~\rf{anroots}, $\vec{\Lambda}_1= 2\vec{v}_1$, which has to be split in two
components. One, in the weight lattice of $A_{n-1}$, and another,
corresponding to the $u(1)$ subalgebra of $\bar{g}_0$, orthogonal to it. The
decomposition is as follows
\be
\vec{\Lambda}_1 \>=\> \vec{\lambda}_{A_{n-1}}\> +\> \vec{\lambda}_{u(1)}
\ee
with
\be
\vec{\lambda}_{A_{n-1}}\>= \> \sum_{i=1}^{n-1} {2(n-i)\over n}\> (\vec{v}_i -
\vec{v}_{i+1}) \>, \qquad
\vec{\lambda}_{u(1)}\> =\> {2\over n}\> \sum_{i=1}^n \vec{v}_i\> =\> {1\over
\sqrt{\pi n}}\> \vec{u}\>.
\ee
This allows one to calculate the required quadratic Casimir
\be
C_2(A_{n-1})\,=\,\langle\,\vec{\lambda}_{A_{n-1}}
\,,\, \vec{\lambda}_{A_{n-1}}\,+\,
2\vec{\d}_{A_{n-1}}\,\rangle \,=\,
\frac{(n-1)(n+2)}{n}\> \vec{\Psi}_{A_{n-1}}^2, 
\lab{casimircn2}
\ee
where we have used 
\be
2\vec{\d}_{A_{n-1}}\,=\,\sum_{i=1}^{[{n}/2]}{(\vec{v}_i
\,-\,\vec{v}_{n+1-i})}(n+1-2i),
\lab{sumroots}
\ee
with $[n/2]$ the integer part of $n/2$. The same
result can be obtained by considering the other representation with highest
weight $\vec{\Lambda}_2$.

Therefore, the SSSG theories associated with the symmetric space $Sp(n)/U(n)$
are integral perturbations of either the WZNW model corresponding to $SU(n)$
at level $k$ plus a massless boson ($p=0$) or a coset CFT of the form
$[SU(n)_k/U(1)^p]\times U(1)$, whose central charge is given by~\rf{centralB}:
\be
c_{\rm CFT}\> =\>{k\>(n^2-1) \over k+n}\> +\> 1\> -\> p\>.
\ee
The conformal dimension of the perturbation is given by~\rf{dimensionfiB}:
\be
\Delta_{\Phi}\,=\,\frac{(n-1)(n+2)}{n(k+n)}+\frac{2}{kn},
\lab{dimencn}
\ee
and there is a different SSSG theory for $0\leq p\leq n-1$. The
perturbation is relevant for $k>2$.

\sect{Integrability of the SSSG theories.}
\ \indent\label{integra}

The classical integrability of the Homogeneous (HSG)
and Symmetric Space (SSSG) sine-Gordon theories is a consequence of the
relationship between their equations of motion and the non-abelian affine 
Toda equations, which admit a zero-curvature representation. This
implies the existence of an infinite number of conserved quantities, whose
construction by means of the Drinfel'd-Sokolov procedure will be summarized 
in the first part of this section.

Concerning the HSG theories, their quantum integrability was
established in~\cite{hsg1} by explicitly checking that the conserved
quantities of scale dimension $\pm2$ remain conserved in the
quantum theory after an appropriate renormalization, and invoking a well known
argument due to Parke~\cite{parke}. 

For the SSSG theories, we expect that
something similar happens and, hence, that they are also quantum integrable.
However, the variety of different types of SSSG theories makes difficult to
thoroughly check this conjecture. In general, and similarly to the HSG models,
most of the SSSG theories exhibit classically conserved quantities of scale
dimension~$\pm1, \pm2, \ldots$, and it should be possible to generalize the
proof in~\cite{hsg1} to show that the conserved quantities of scale
dimension
$\pm2$ also give rise to quantum conserved quantities. Nevertheless, there
is an exception to this pattern: the SSSG theories associated with symmetric
spaces of maximal rank, {\it i.e.\/}, with symmetric spaces $G/G_0$ whose
rank is ${\rm rank\/}(G/G_0) = {\rm rank\/}(G)$, which only have conserved
quantities of odd scale dimension $\pm1, \pm3, \ldots$. According
to~\rf{dim}, they are relevant perturbations of either the WZNW model
corresponding to $G_0$ ($p=0$) or a coset CFT of the form $G_0/U(1)^p$ with
$0\leq p\leq {\rm rank\/}(G_0)$. The type~I symmetric spaces of maximal 
rank are the following~\cite{helgas}:
\bea
&&SO(2n)/SO(n)\times SO(n)\>, \quad SO(2n+1)/SO(n)\times SO(n+1)\>, \nn
\noalign{\vskip 0.2truecm}
&&SU(n)/SO(n) \>, \quad Sp(n)/U(n) \>, \quad E_6/Sp(4) \>, \quad E_7/SU(8) \>,
\nn 
\noalign{\vskip 0.2truecm}
&&E_8/SO(16) \>, \quad F_4/Sp(3)\times SU(2) \>, \quad G_2/SU(2)\times
SU(2)\>. 
\lab{SplitSS}
\ena
Notice that there is one for each simple compact Lie group $G$, which is
related to its (unique) maximally non compact real form, also known
as `split form'. Besides $SU(2)/SO(2)$, which corresponds to the
sine-Gordon theory, the simplest symmetric space of maximal rank is
$SU(3)/SO(3)$. Then, eq.~\rf{dim} reads $0\leq p\leq1$, which means that it
gives rise to two different SSSG theories, depending on the choice of
$\Lambda_\pm$. They are just the integrable perturbations of the $SU(2)_k$
WZNW model $(p=0)$ and its $\widetilde{so(2)}$ reduction $(p=1)$
constructed by Brazhnikov~\cite{brazhnikov}, which is identified with the
perturbation of the usual $\ZZ_k$-parafermions by the second thermal
operator.
 
In this section we explicitly prove the quantum integrability
of the SSSG theories associated with a maximal rank symmetric space and
$p=0$, which will be called `Split models'. Actually, they can be distinguished
as the only ones that provide relevant perturbations of WZNW models. Namely, we will check
that their simplest higher spin classically conserved densities, which have spin
$\pm3$ instead of $\pm2$, remain conserved in the quantum theory after an
appropriate renormalization. According to~\cite{parke}, this is enough to
establish their quantum integrability, which, together with the results
of~\cite{hsg1} for the HSG theories, supports the conjecture that all the SSSG
theories are quantum integrable.

\subsection{Classical integrability.}
\indent\ \

The zero-curvature form of the equations of motion~\rf{equation12} is
\be
[\,{\partial}_{+}\,+\,m\> h{\Lambda}_{+}h^{\dag}
\,,\,{\partial}_{-}\,+\,m\> \Lambda_{-}\,-\,
{\partial}_{-}h h^{\dag}]=0\>.
\lab{zerocur}
\ee
while the constraints arise naturally in the group theoretical
description of the non-abelian affine Toda equations~\cite{nt,Luiz}. In order to
use the generalized Drinfel'd-Sokolov construction of~\cite{drinfel} to get
the infinite number of conserved densities of~\rf{zerocur}, we have to
associate the zero-curvature equation with a loop algebra. In our case, the
relevant loop algebra is
$\cl{(g,\sigma)}
\,=\, \bigoplus_{j\in\ZZ} {\cal L\/}(g,\sigma)_j$ with
\be
{\cal L\/}(g,\sigma)_{2j}\> =\> \lambda^j \otimes g_0 \>, \quad {\rm and}\quad
{\cal L\/}(g,\sigma)_{2j+1} \>=\> \lambda^j \otimes g_1 \>,
\lab{loop}
\ee
where $\lambda$ is a spectral parameter; if $\sigma= \sigma_{(\vec{s};r)}$,
then $\cl{(g,\sigma)}$ is related to the (twisted if $r\not=1$) affine
Kac-Moody algebra $\bar{g}^{(r)}$ without central extension. Moreover, since 
\be 
[{\cal L\/}(g,\sigma)_j\>, \> {\cal L\/}(g,\sigma)_k]\subset {\cal
L\/}(g,\sigma)_{j+k}\>,
\ee
the subspaces~\rf{loop} define an integer gradation of $\cl{(g,\sigma)}$.
The equations of motion~\rf{zerocur} remain unchanged under the transformation
$\Lambda_{\pm}{\mapsto}{\l^{\mp{1}}}\,{\otimes}\,\Lambda_{\pm}$ and, hence, the
zero-curvature equation can actually  be associated with $\cl(g,{\sigma})$
and the Lax operator $L=\partial_{-}\,+\,\Lambda\,+\,q$, where
\be
\Lambda \>=\>  m\> \lambda \otimes \Lambda_{-} \,\in
\,{\cl(g,{\sigma})}_1
\quad {\rm and}\quad q\,=\,-\partial_{-}h
 h^{\dag}\,\in \,{\cl(g,{\sigma})}_0.
\lab{lax1}
\ee

The generalized Drinfel'd-Sokolov construction goes as follows~\cite{drinfel}.
First, there is some local function~$y$ of the `potential'~$q$ of the
form~\footnote{By a local function of $q$ we mean a $\partial_-$-differential
polynomial in the components of $q$, {\it i.e.\/}, a polynomial in the
components of $q$, $\partial_- q$, $\partial_-^2 q$, $\ldots$}
\be
y\,=\,\sum_{n>0} y^{a}(n) \lambda^{-k} \otimes t^{a} \in 
{Im(ad_\Lambda)}_{<0},
\lab{functional}
\ee
that `abelianizes' the Lax operator in the following sense:
\be
e^y \>L\> e^{-y}\,=\,e^y
(\partial_{-}\,+\,\Lambda\,+\,q) e^{-y}\,=\,
\partial_{-}\,+\,\Lambda\,+\,H\>, 
\lab{abel}
\ee
where $H \,\in \,{\rm Ker\/}({\rm ad\/}_\Lambda)_{ \leq 0}$ is another local 
function of~$q$, and $\{t^a\}$ is a basis for $g$ whose standard realization is
presented in the Appendix. Then, eqs.~\rf{zerocur} and~\rf{abel} imply
\be
e^y \Bigl(\partial_{+}\,+\,mh(\lambda ^ {-1} 
\otimes \Lambda_{+})h^{\dag}\Bigr) e^{-y}\,=\,
\partial_{+}\,+\,\overline{H}.
\lab{abel1}
\ee
where $\overline{H}$ also takes values in  ${\rm Ker\/}({\rm ad\/}_\Lambda)_{
\leq 0}$ and, therefore, the zero-curvature equation becomes
\be
\partial_{-}\overline{H}-\partial_{+}{H}\,=\,[\,\overline{H},{H} \,]\>.
\lab{zerocur1}
\ee

The components of this equation along the centre of ${\rm
Ker\/}({\rm ad\/}_\Lambda)$ provide an infinite number of local conservation
laws. To be precise, let $b \in {\rm Cent\/}\bigl({\rm
Ker\/}({\rm ad\/}_\Lambda)\bigr)_j$ with $j>0$, and define
\be
{\cal I}_{j}^{(0)}[b] = \langle b\>, \> H\rangle  \>, \qquad
\overline{{\cal I}}_{j}^{(0)}[b] = \langle b\>, \> \overline{H}\rangle \>.
\lab{Cdensity}
\ee
The corresponding local conservation law is
\be
\partial_+ {\cal I}_{j}^{(0)}[b]\>= \> \partial_- \overline{{\cal
I}}_{j}^{(0)}[b]\>.
\lab{Claw}
\ee
Moreover, one can check that the conserved densities given by~\rf{Cdensity} have
scale dimension $j+1>1$ with respect to the scale transformations
$x_{\pm}\rightarrow {x_{\pm}}/\rho$, which means that they give rise to
conserved quantities of scale dimension~$j>0$. The same procedure can be
repeated by changing the Lax operator $L$ by 
\be
\overline{L}\> =\> \partial_+ \> +\> m\lambda^{-1}\otimes \Lambda_+ \> +\>
h^\dagger\partial_+ h 
\lab{Lbar}
\ee 
in eq.~\rf{abel}, and the result is the construction of another infinite number
of local conserved quantities with scale dimension $j< 0$, {\it i.e.\/},
negative scale dimension. Notice that $L$ and $\overline{L}$ are conjugate by the
transformation~\rf{parity} and, moreover, that the dimensions of ${\cal
L\/}(g,\sigma)_j$ and ${\cal L\/}(g,\sigma)_{-j}$ are equal. Therefore, for
each conserved quantity of positive scale dimension $j$ there will be another
conserved quantity of scale dimension $-j$, and both are conjugate
by~\rf{parity}. In particular, if the theory is parity invariant both conserved
quantities are parity conjugate. This allows one to restrict the analysis to
the conserved quantities with positive scale dimension. 

Taking into account all this, the resulting number of classically conserved
quantities with scale dimension $\pm j$ is given by the dimension of ${\rm
Cent\/}\bigl({\rm Ker\/}({\rm ad\/}_\Lambda)\bigr)_{\pm j}$. These dimensions
can be easily calculated when $\Lambda_\pm$ are chosen to be regular, which
means that ${\rm Ker\/}({\rm ad\/}_{\Lambda_\pm})$ is a Cartan subalgebra of $g$
and  $p= {\rm rank\/}(G)-{\rm rank\/}(G/G_0)$ in~\rf{dim}. Then, the
Drinfel'd-Sokolov construction produces exactly ${\rm rank\/}(G/G_0)$ local
conserved quantities for each odd scale dimension $\pm1, \pm3, \ldots$, and
${\rm rank\/}(G)-{\rm rank\/}(G/G_0)$ for each even scale dimension $\pm2,
\pm4, \ldots$. 

Consider now a SSSG theory related to a symmetric space of maximal rank
with $p$ in the range~\rf{dim} ($p=0$ corresponds to the case when
$\Lambda_\pm$ are regular). Then, there  are ${\rm rank\/}(G/G_0)-p$ local
conserved quantities for each odd scale dimension, $\pm1, \pm3, \ldots$,
and no conserved quantities with even scale dimension. This can be easily
proved by choosing the Cartan subalgebra such that it contains the abelian
subalgebra $g_0^0 = u(1)^p$ given by~\rf{lambcond}.

These results agree with what has been anticipated at the
beginning of this section: in general, the simplest higher spin conserved
quantities of a SSSG theory have spin $\pm2$. However, if ${\rm
rank\/}(G/G_0)={\rm rank\/}(G)$, the simplest conserved quantities will have
scale dimension $\pm3$.

\subsection{Quantum integrability of the Split models.}

\indent \ \
In the following, we will check that, after a suitable renormalization, the
conserved densities of spin $\pm1$ and $\pm3$ remain conserved in the quantum
version of the Split models. The explicit expressions for the relevant
classical conserved densities are given in the Appendix. The proof will be
similar to the one presented in~\cite{hsg1} for the HSG theories, which uses
conformal perturbation theory. However, since the simplest higher spin conserved
quantity is of spin $3$ instead of $2$, this case will be even more involved, and
in order to avoid unnecessary complications we will restrict ourselves to the
Split models with $G_0$ simple. We will also fix the normalization of the
invariant bilinear form of $g$, $\langle\>, \>\rangle$, such that
$\vec{\Psi}_{g_0}^2 =2$.

Since the quantum SSSG
theories can be described as perturbed conformal field theories, the existence
of quantum conserved quantities can be investigated by using the methods
of~\cite{pertcft}. In the presence of the perturbation~\rf{pertur} any
chiral field ${\cal I}(z)$, which in the unperturbed CFT satisfies
$\bar\partial {\cal I}(z)=0$, acquires a $\bar{z}$ dependence given by
\be
\bar{\partial}{\cal I} (z,\bar{z})\,=\,-k{m^2}\oint_z
\frac{dw}{2 \pi i}
\Phi(w,\bar{z}){\cal I} (z)\>,
\lab{zamo1}
\ee
where we have introduced the notation $z=x_-$, $\bar{z} = x_+$, $\partial=
\partial_-$, and $\bar{\partial}= \partial_+$ reminiscent of euclidean
space. This  contribution actually corresponds to the lowest order in
perturbation theory; however, if the condition of super-renormalizability
at first order $2 \Delta_{\Phi}
\leq 1$ is satisfied,  no counterterms are needed to
renormalize~\rf{pertur}, and the previous equation is expected to be
exact~\cite{pertcft,cardy}. Actually, for any SSSG theory, this condition
is always fulfilled for $k$ above some minimal value characteristic of the
theory (see the comments at the end of Section~\ref{typeI1}). Therefore, in
the perturbed CFT the chiral field ${\cal I}(z)$ will become a conserved
quantity if the right-hand-side of~\rf{zamo1} is a total
$\partial$~derivative, i.e. if~\rf{zamo1} can be written as
$\bar{\partial}{\cal I}= \partial\bar{\cal I}$ where $\bar{\cal I}$ is
another field of the original CFT. For this condition to be satisfied, the
residue of the simple pole in the OPE between ${\cal I}(z)$ and
$\Phi(w,\bar{z})$ has to be a total $\partial$~derivative; namely,
\be 
{\Phi}(w,\bar{z}) {\cal I} (z) \,=\,\sum_{n>1} \,\frac{\{ {\Phi}  {\cal I}
\}_n(z,\bar{z})}{{(w-z)}^n}\,+\,
\frac{\partial\bar{\cal I}(z,\bar{z})}{(w-z)}\,+\cdots
\lab{ope}
\ee
Notice that the residues of the simple poles in the OPE's ${\Phi}(w,\bar{z})
{\cal I} (z)$  and ${\cal I} (z){\Phi}(w,\bar{w})$  differ only in a total
$\partial$~derivative  and, in practice, we will always consider the latter 
whose expression is usually simpler.

For the Split models, the unperturbed CFT is just the WZNW model
corresponding to $G_0$ at level $k$. Using the conventions of the
Appendix, the subset of generators of $g$ given by $t^\alpha$ for each
positive root $\vec{\alpha}$ of $g$ provides a suitable basis of generators
for the Lie subalgebra $g_0$. This means that the operator algebra of the
$G_0$--WZNW model can be realized as a subset of the operator algebra of the
WZNW model associated to~$G$ that, in particular, includes the chiral
currents
$J^\alpha(z)$ and $\bar{J}^\alpha(\bar{z})$ which satisfy the OPE
\be
J^{\alpha}(w)J^{\beta}(z)\,=\,\frac{\hbar^2 k\> \d^{\alpha \beta}}{{(w-z)}^2}
\,+\,\frac{\hbar\> f^{\alpha \beta \gamma}\, J^{\gamma}(z)}{(w - z)}\,+\,\cdots
\lab{ope1}
\ee
Moreover, it will be useful to recall the following well known
identities~\cite{wzw,olive}. The first one is 
\be
J^\alpha(z)\> h(w,\bar{w}) \>=\> -\>\hbar\> {t^\alpha\> h(w,\bar{w})\over z-w} \>
+\cdots\>,
\lab{HPrimary}
\ee
which is satisfied in an arbitrary representation of $G_0$ and exhibits that the
WZNW field $h$ is a primary field. The second is the relation between the WZNW
field and the chiral currents
\be
\hbar\>(k\>+ \> h_{g_0}^\vee) \> \partial h\> =\> \sum_\alpha \bigl(J^\alpha\>
t^\alpha \> h)\>,
\lab{Null}
\ee
where $(AB)(z)$ is the normal ordered product of two operators $A(z)$ and
$B(z)$, which will be defined by adopting the conventions of~\cite{baisC}.
At this point, it is also convenient to recall that
the classical expressions are recovered from the quantum ones by means of
(see~\rf{lax1} and~\rf{Null})
\be 
J^\alpha t^\alpha \> =\> -\> (\hbar k)\> q\>, \qquad (\hbar k) \rightarrow
\frac{1}{\b^2}\>, \quad
 {\rm and} \quad k  \rightarrow \infty\>,
\lab{equiv}
\ee
which, according to~\rf{level}, amounts to take the classical ($\hbar\ra 0$)
or weak coupling ($\beta^2 \ra0$) limits. Therefore, the differential
polynomials in the components of $q$ that appear in the classical local
conserved densities have to be changed into differential polynomials in the
chiral currents $J^\alpha$. In the rest of this section we
will show explicitly the Plank constant to make the distinction between
classical and quantum contributions simpler.

\subsubsection{Quantum spin-2 conserved densities (spin-1 conserved
quantities).}

\indent \ \

Consider a generic normal ordered local spin-2 operator
\be
{\cal I}_2(z)\,=\,D_{\a\b} ({J^\a}{J^\b})(z),
\lab{spin2}
\ee
where the numerical coefficients satisfy $D_{\a\b}=D_{\b\a}$.
Taking into account that the perturbing operator $\Phi =\langle \Lambda_-,
P\rangle$, with $P= (h\Lambda_+ h^\dagger)$, is just the WZNW field taken in
the representation of $G_0$ provided by $[g_0, g_1]\subset g_1$,
eqs.~\rf{HPrimary} and~\rf{Null} become
\bea
&& J^\alpha(z)\> \langle v\>,\> P(w,\bar{w}) \rangle \>=\>
-\> \hbar\>\frac{\langle[v\> ,\> t^\alpha]
\> ,\> P(w,\bar{w})\rangle}{(z-w)}\> +\> \cdots \\
\noalign{\vskip 0.2truecm} 
&& \hbar\>(k+h_{g_0}^{\lor})\,\partial\langle v\>,\> P \rangle \> =\> 
 \bigl(J^\alpha\>\langle [v\>, \>t^\alpha]\>, \>  P
\rangle\bigr)\>, 
\lab{spin22}
\ena
where $v$ is an arbitrary element in $g_1$. Then, it is easy to show
that the residue of the simple pole in the OPE ${\cal I}_2 (z) \Phi(w,\bar{w})$
is given by
\be
{\rm Res}
\Big( {\cal I}_2 (z)\> \Phi(w,\bar{w})\Big)\> =\> -2 \>\hbar\> 
\bigl(J^\b\>
\langle D_{\a\b}\,[\,\Lambda_{-},t^\a\,] , P(w,\bar{w})
\,\rangle\bigr)\>.
\lab{res2}
\ee
Taking into account~\rf{spin22}, this residue is a total 
$\partial$~derivative if the tensor $D_{\a\b}$
satisfies~\footnote{Explicit factors of $\hbar k$ are included to take
account of the relation $J^\alpha t^\alpha = -(\hbar k)\> q$ in the
$k\ra
\infty$ limit and recover the classical expressions given in  the
Appendix (see eq.~\rf{equiv}).}
\be
(\hbar k)^2
D_{\a\b}\,[\,\Lambda_{-},t^\b\,]\>
=\> \frac{1}{2}\> [\vec{\m}{\cdot}\vec{t}\> ,\> t^\a\,],
\lab{tensorspin2}
\ee
for any ${\rm rank\/}(g)$-component vector $\vec\m$. This leads to the following
solutions labelled by $\vec\m$ ($\vec{\l}$ is defined in~\rf{Lambase})
\be
(\hbar k)^2 D_{\a\b}({\vec{\m}})\> =\> \frac{1}{2}\>
\frac{\vec{\m}{\cdot}\vec{\a}}{\vec{\lambda}{\cdot}\vec{\a}}\> 
\d_{\a\b}\> =\> m\> D_{\a\b}^{(0)}({\vec{\m}})\>,
\lab{tensor2form}
\ee
which satisfy
\be
{\rm Res} \Big( {\cal I}_2 (z)\> \Phi(w,\bar{w})\Big)\> =\>
-\> \frac{(k+h_{g_0}^{\lor})}{{k}^2}
 \> \partial \langle\,{\vec{\m}{\cdot}\vec{t}}\> ,\> P \rangle (w,\bar{w})\>.
\lab{conserv2}
\ee
Therefore, the quantities
\be
(\hbar k)^2 {\cal I}_2({\vec{\m}})\,=\,
\frac{1}{2} \sum_{\a>0}\,\frac{\vec{\m}{\cdot}\vec{\a}}
{{\vec{\lambda}}{\cdot}{\vec{\a}}}\> ({J^\a}{J^\a}) 
\lab{i2}
\ee
provide ${\rm rank\/}(g)$ linearly independent conserved densities of spin-2. The
field $\bar{\cal I}_2(\vec{\m})$ that satisfies the conservation law
$\bar{\partial}{\cal I}_2 = \partial\bar{\cal I}_2$ is obtained through the
explicit evaluation of the integral~\rf{zamo1}; the result is
\be
(\hbar k)^2 \bar{\cal I}_{2}({\vec{\m}})\> =\>
-m^2\> (\hbar k)^2\> \Bigg(1+ \frac{4 h_{g_0}^{\lor}-\Psi_{g}^{2} 
h_{g}^{\lor}}{4k}\Bigg) \> 
\langle\vec{\m}\cdot \vec{t}\> ,\> P\rangle \>.
\lab{i2bar}
\ee

It has already been pointed out that these quantum conserved quantities can be
understood as the renormalization of the classical ones calculated in the
Appendix. In this case, the relationship is particularly simple: the spin-2
quantum conserved densities equal the normal ordered classical conserved
ones of scale dimension~2, up to the multiplicative renormalization
in~\rf{i2bar}. In more general cases, one has to add quantum
$O(1/k)\simeq O(\hbar\beta^2)$ corrections; however, the number of them must be
finite as a reflection of the fact that the perturbation is relevant and,
therefore, the quantum theory is super-renormalizable. 

As expected, the stress-energy tensor is a
particular example of a spin-2 conserved quantity. It is recovered
from~\rf{i2} for the particular choice $\vec\m=\vec\l$: 
\be
(\hbar k)^2{\cal I}_2({\vec{\lambda}})\,=\,
\frac{1}{2}\,
({J^\a}{J^\a}) \,=\,-\> (k+h_{g_0}^{\lor})\> \Bigl\lbrace \frac{-1}
{2(k+h_{g_0}^{\lor})}\> ({J^\a}{J^\a})\Bigr\rbrace 
\> =\> -\> (k+h_{g_0}^{\lor})\> T_{z,z} \>,
\lab{suga1}
\ee
in agreement with the Sugawara construction. Consequently,
\be
(\hbar k)^2 \bar{\cal I}_{2}({\vec{\lambda}})\,
=\,-{m^2}\> (\hbar k)^2  
\,\Bigg(1+\frac{4 h_{g_0}^{\lor}-\Psi_{g}^{2} 
h_{g}^{\lor}}{4k}\Bigg)\> \langle \Lambda_{-} \>,\>  P\rangle\, = \,-\>
(k+h_{g_0}^{\lor})\> T_{z,\bar{z}}\>.
\lab{suga2}
\ee

\subsubsection{Quantum spin-4 conserved densities (spin-3 conserved 
quantities).}

\indent \ \

Up to total $\partial$-derivatives, the most general form of
a spin-4 normal ordered operator
constructed with the WZNW currents $J^\alpha$ is
\be
{\cal I}_{4}{(z)}=R_{\a\b\g\rho } {(J ^{\a}(J^{\b}(J^{\g}J^{\rho})))\,(z)} +
P_{\a\b\g} (J ^{\a}(J^{\b}{\partial J^{\g}})) \,(z) +
Q_{\a\b} (J ^{\a} {{{\partial}^{2}}J^{\b}})\,(z)\> ,
\lab{spin4}
\ee
with the following constraints:
\bea
&&R_{\a\b\g\rho } \,= \, R_{(\a\b\g\rho)},\qquad Q_{\a\b} \,= \,Q_{\b\a}\>,\nn
&&P_{\a\b\g}\,= \,P_{\b\a\g}, \qquad
P_{\a\b\g}\,+\,P_{\g\b\a}\,+\,P_{\a\g\b}\,= \,0\>, 
\lab{const12}
\ena
where the first one indicates that $R_{\a\b\g\rho }$ is a totally symmetric
tensor. Then, the residue of the simple pole in the OPE
${\cal I}_4 (z) \Phi(w,\bar{w})$ is
\bea
{\rm Res}
 \bigg( {\cal I}_{4}(z)  \Phi (w,\bar{w}) \bigg) &=& \hbar\> \bigg\{
(J^{\a}(J^{\b}(J^{\g} {\langle}\, \Omega _{\a \b\g}\, ,\, P \,{\rangle} )))+ (
\partial J^{\a}(J^{\b} {\langle}\, M_{\a \b}\, ,\, P\,{\rangle})) +{}
\nonumber\\
&& {}+ (\partial ^2 {J^{\a}}{\langle}\, T_ {\a} 
\, ,\, P \,{\rangle}) \bigg\}(w,\bar{w}),
\lab{res4}
\ena
where
\bea
&& {\Omega_{\a \b\g} }\,=\, 4\, R_{\a\b\g\rho} 
\,[{\Lambda }_ {-}  , { t^{\rho} }],
\lab{ten1}\\
\noalign{\vskip 0.4truecm}
&& {M_{\a \b}}\,=\,-12\,{ \hbar} \,R_{\a\b\g\rho}\,
[\,[{\Lambda }_ {-}  ,  t^{\g} ] ,  t^{\rho} ] 
\, - 2\, (P_{\a\b\g}-P_{\b\g\a})\,[{\Lambda }_ {-}  , t^{\g} ],
\lab{ten2}\\
\noalign{\vskip 0.4truecm}
&& {T_ {\a}}\,=\,2\,{{ \hbar}^{2}}\,R_{\a\b\g\rho}\,
[\,[\,[{\Lambda }_ {-}  ,t^{\b} ] t^{\g} ] ,  t^{\rho} ]
 \,+\,{ \hbar} (P_{\a\b\g}-P_{\g\b\a})\,[\,[{\Lambda }_ {-} 
 ,  t^{\b} ] ,  t^{\g} ] + \nn
\noalign{\vskip 0.2truecm}
&&\qquad \qquad +\> 
(2\,Q_{\a\rho}\,+\,\hbar f^{\b\g\a}P_{\b\rho\g})
\,[{\Lambda }_ {-}  , { t^{\rho} }].
\lab{ten3}
\ena

The condition that ${\cal I}_4(z)$ is a conserved quantity
is equivalent to the existence of two 
tensors $F_{\a\b}$ and $R_{\a}$ taking values in $g_1$ such that
\be
Res \bigg( I_{4}(z)  \Phi(w,\bar{w}) \bigg) \,=\,\partial{(J^{\a}(J^{\b}
{\langle}\, F_{\a \b}\, ,\, P\,{\rangle}))}\,+
\,\partial{(\partial{J^{\a}}{\langle}\, R_{\a}\, ,\, P\,{\rangle})}\>.
\lab{geni4}
\ee
We will assume that $F_{\a\b}$ is totally
symmetric because, otherwise, its antisymmetric part $F_{[\g\b]}$ 
can be absorbed in $R_\a$ through the transformation
\be
R_{\a}
\,\ra\,R_{\a}\,+\, \frac {\hbar}{2}f^{\a\b\g} F_{[\g\b]}\>.
\ee
If we define
\be
\widehat{F}_{\a\b}\,=\,\frac{F_{\a\b}}{\hbar(k+h_{g_o}^{\lor})}
\quad {\rm and}\quad \widehat{R}_{\a}\,=\,\frac{R_{\a}}
{\hbar(k+h_{g_o}^{\lor})}\>,
\lab{hats}
\ee
eqs.~\rf{ten1}--\rf{hats} can be combined to obtain the following tensor
relations
\bea
&& 12\, R_{\a\b\g\rho} \,[{\Lambda }_ {-} 
, { t^{\rho} }] \,=\,
[\,\widehat{F}_{\a \b}\,,\,t^{\g}\,] \,+\,[\,\widehat{F}_{\a
\g}\,,\,t^{\b}\,]\,+\,[\,\widehat{F}_{\g\b}\,,\,t^{\a}\,],
\lab{tensorR}\\
\noalign{\vskip 0.4truecm}
&&- 2\, (P_{\a\b\g}-P_{\b\g\a})[{\Lambda }_ {-}  , t^{\g}\, ] \,
\,=\,2\,F_{\a\b}\,+\,[\widehat{R}_{\a},t^{\b}]\,+\,\hbar\bigg\{ \,[\widehat{F}_{\b\rho},[t^{\a},t^{\rho}]\,]\,
+\nn
\noalign{\vskip 0.2truecm}
&& \qquad\qquad + \,[\,[\widehat{F}_{\a\b},t^{\rho}],t^{\rho}] \,+\,
[\,[\widehat{F}_{\a\rho},t^{\b}],t^{\rho}] +
\,[\,[\widehat{F}_{\b\rho},t^{\a}],t^{\rho}]\, \bigg\},
\lab{tensorP} \\
\noalign{\vskip 0.4truecm}
&& -2\,Q_{\alpha\beta}\,[\,{\Lambda }_{-},{ t^{\b}}\,] = -{R}_{\a}
\,+\, \hbar \bigg\{ \,(P_{\a\b\g}-P_{\g\b\a})\,[\,[{\Lambda }_ {-}
  , t^{\b} \,], t^{\g}]\, + \,
{f ^{\b\g\a}}\,P_{\b\rho\g}\,\,[{\Lambda }_ {-},t^{\rho}\, ] \bigg\}
+\nn
\noalign{\vskip 0.2truecm}
&& \qquad\qquad + 2{\hbar}^{2} \bigg\{ \,{R_{\a\b\g\rho}}\,[\,
[\,[{\Lambda }_ {-}  , t^{\b}] ,t^{\g}],{ t^{\rho}}]\, +
\,{\frac{1}{3}}\,{f ^{\xi\g\rho}}{f ^{\b\rho\a}}\,
[\,\widehat{F}_{\b\xi},t^{\g}\,]\bigg\}\>,
\lab{tensorQ}
\ena
where~\rf{tensorR} does not contain
any explicit quantum correction at all, which means that the same relation
will hold at the classical level. In fact, taking into account the
explicit expressions for the classical densities given in the
Appendix and~\rf{equiv}, for each ${\rm rank\/}(g)$-component vector $\vec\mu$
there is a classical solution of~\rf{tensorR}--\rf{tensorQ} given by
\be
 R_{\a\b\g\rho}= \frac{(\hbar k)^4}{m^3}\,R_{\a\b\g\rho}^{(0)}(\vec{\m})\>,
\quad P_{\a\b\g}= -\frac{(\hbar k)^3}{m^3}\, P_{\a\b\g}^{(0)}(\vec{\m})
\>, \quad Q_{\alpha\beta} = \frac{(\hbar
k)^2}{m^3}\, Q_{\alpha\beta}^{(0)}(\vec{\m})\>,
\lab{classquan}
\ee
which satisfy the classical limit of those equations: 
\bea
&&12 \,{m^3} R_{\a\b\g\rho}^{(0)}(\vec{\m})\,[{\Lambda }_ {-} ,{t^{\rho}}]  
= {{(\hbar k)}^4} \bigg\{ [\widehat{F}_{\a \b}^{(0)}(\vec{\m}) , t^{\g}]
 + [\widehat{F}_{\a \g}^{(0)}(\vec{\m}) , t^{\b}]+
 [\widehat{F}_{\b \g}^{(0)}(\vec{\m}), t^{\a}] \bigg\},
\lab{classR} \\
\noalign{\vskip 0.2truecm}
&&2\,{m^3}\, \bigl(P_{\a\b\g}^{(0)}(\vec{\m})-P_{\b\g\a}^{(0)}(\vec{\m})\bigr)\,
[{\Lambda }_ {-},t^{\g} ] ={{(\hbar k)}^3} \,
\bigg\{ 2\,(\hbar k)\,\widehat{F}_{\a\b}^{(0)}(\vec{\m})\,+\,
[\widehat{R}_{\a}^{(0)}(\vec{\m}),t^{\b}] \bigg\},
\lab{classP}\\
\noalign{\vskip 0.2truecm}
&&2\,{m^3}\, Q_{\alpha\beta}^{(0)}(\vec{\m})\,[{\Lambda }_ {-}  , 
{ t^{\b} }]\,= - {{(\hbar k)}^3} {\widehat{R}}_{\a}^{(0)}(\vec{\m}).
\lab{classQ}
\ena
Using eqs.~\rf{tensor1}--\rf{tensor3}, one can obtain the following
explicit expressions for the classical limit of the corresponding tensors
$\widehat{F}_{\a\b}(\vec{\m})$ and $\widehat{R}_{\a}(\vec{\m})$
\bea
&&{(\hbar k)}^3 R_{\a}^{(0)}(\vec{\m})\,=\,
-\frac{\vec{\a}\cdot\vec{\m}}{(\vec{\lambda}\cdot\vec{\a})^2}\,\,t^{\bar{\a}}\>,
\lab{classR2}\\
\noalign{\vskip 0.4truecm}
&&{{(\hbar k)}^4} F_{\a\b}^{(0)}(\vec{\m})\,=\,\frac{t^{\bar{\g}}}
{4}
\Bigg\{\frac{1}{4}\Bigg(
\frac{\vec{\a}\cdot\vec{\m}}{{(\vec{\lambda}\cdot\vec{\a})}^2}
f^{\bar{\a}\b\bar{\g}}\,+\,\frac{\vec{\b}\cdot\vec{\m}}
{{(\vec{\lambda}\cdot\vec{\b})}^2}f^{\bar{\b}\a\bar{\g}}\Bigg)+{}
\nn
\noalign{\vskip 0.2truecm}
&&\qquad\qquad +\>
\Bigg(\frac{f^{{\g}\a{\b}}}{\vec{\lambda}\cdot\vec{\g}}\,+\,
\frac{3}{4}\frac{\vec{\lambda}\cdot\vec{\g}\>
f^{\bar{\a}\g\bar{\b}}}{(\vec{\lambda}\cdot\vec{\a})
(\vec{\lambda}\cdot\vec{\b})}\Bigg)
\Bigg(\frac{\vec{\b}\cdot\vec{\m}}{\vec{\lambda}\cdot\vec{\b}}
\,-\,\frac{\vec{\a}\cdot\vec{\m}}{\vec{\lambda}\cdot\vec{\a}}
\Bigg)-{}
\nn
\noalign{\vskip 0.2truecm}
&&\qquad\qquad -\> 
\frac{\vec{\g}\cdot\vec{\m}}{4\> \vec{\lambda}
\cdot\vec{\g}}\Bigg(
\frac{f^{\bar{\b}\a\bar{\g}}}{\vec{\lambda}\cdot\vec{\b}}
\,+\,\frac{f^{\bar{\a}\b\bar{\g}}}{\vec{\lambda}\cdot\vec{\a}}
\Bigg)\Bigg\}-
\frac{(\vec{\a}\cdot\vec{\m})\> \d_{\a\b}}{2\>
{(\vec{\lambda}\cdot\vec{\a})}^2}\>\vec\a\cdot \vec{t}\>, 
\lab{classF}
\ena
where $\vec\a\cdot\vec{t}=\a^A{t}^A$ is in the Cartan subalgebra of~$g$. 
In particular, when $\vec\m=\vec\l$ the previous
equations reduce to
\bea
&&{{(\hbar k)}^3}\widehat{R}_{\a}^{(0)}(\vec{\l})
=-\frac{1}
{\vec{\l}\cdot\vec{\a}}\, t^{\bar\a},
\lab{classRR}\\
\noalign{\vskip 0.2truecm}
&&{(\hbar k)^4}\widehat{F}_{\a \b}^{(0)}(\vec{\l})
=-\frac{\d_{\a\b}}{2
\,{\vec{\l}}\cdot \vec{\a}}\>\vec\a \cdot\vec{t} \>.   
\lab{classFF}
\ena

Once we have solved the eqs.~\rf{tensorR}--\rf{tensorQ} in the classical limit,
we will try to find the solutions for the full equations. Since they
become quite complicated, we will only obtain a single one, which is enough
to establish the quantum integrability of these models. Projecting
eqs.~\rf{tensorR}--\rf{tensorQ} on the CSA we get
\bea
&&{\widehat{F}_{\a\b}}^{\bar\g}{{\g}^A}\,+
\,{\widehat{F}_{\b\g}}^{\bar\a}{{\a}^A}\,+
\,{\widehat{F}_{\a\g}}^{\bar\b}{{\b}^A}\,=0,
\lab{tensorRCSA}\\
\noalign{\vskip 0.4truecm}
&&2\,\hbar\,(k+h_{g_o}^{\lor}-\,{\frac{\Psi_{g}^{2} 
h_{g}^{\lor}}{4}})\,{\widehat{F}_{\a\b}}^{A}\,-\,
2\,\hbar\,({\widehat{F}_{\a\b}}^{B} \a^B)\a^A\,
= \,{{\widehat{R}_{\a}}^{\bar\b}}{{\b}^A}\,+{} \nn
\noalign{\vskip 0.2truecm}
&&\qquad\qquad +{}
\,\hbar\,{\{}f ^{\a\rho\d}\,\widehat{F}_{\b\rho}^{\bar\d}
\,+ \,
f ^{\bar\g\a\bar\d}\,{\widehat{F}_{\b\d}}^{\bar\g}\,+\,
f^{\bar\g\b\bar\d}\,{\widehat{F}_{\a\d}}^{\bar\g}{\}}\d^A,
\lab{tensorPCSA} \\
\noalign{\vskip 0.4truecm}
&&\hbar(k+h_{g_0}^{\lor})\widehat{R}_{\a}^A=
\frac{\hbar}{2}\widehat{R}_{\b}^{\bar{\xi}}
f^{\bar{\xi}{\a}\bar{\b}}{\b}^A+\frac{{\hbar}^{2}}{2}
\Bigg\{
f^{\a\rho\d}\Big\{\widehat{F}_{\b\rho}^{\bar{\xi}}
f^{\bar{\xi}\d\bar{\b}}+(\widehat{F}_{\b\rho}^B  {\d^B}) 
\d_{\bar{\d}\bar{\b}}\Big\}+{}
\nn
\noalign{\vskip 0.2truecm}
&&\qquad\qquad +{}
f^{\b\rho\d}
f^{\bar{\xi}\d\bar{\b}}\widehat{F}_{\a\rho}^{\bar{\xi}}
+2f^{\bar{\xi}\rho\bar{\d}}f^{\bar{\d}\rho\bar{\b}}
\widehat{F}_{\a\b}^{\bar{\xi}}\,+\,
2f^{\bar{\xi}\b\bar{\d}}f^{\bar{\d}\rho\bar{\b}}
\widehat{F}_{\a\rho}^{\bar{\xi}}\,+\,
2f^{\bar{\xi}\a\bar{\d}}f^{\bar{\d}\rho\bar{\b}}
\widehat{F}_{\b\rho}^{\bar{\xi}}+{}
\nn
\noalign{\vskip 0.2truecm}
&&\qquad\qquad+{}
2f^{\bar{\a}\rho\bar{\b}}(\widehat{F}_{\b\rho}^B  {\a^B})
\Bigg\}{\b}^A
+2{\hbar}^{2}f^{\bar {\rho}\g\bar{\b}}
\Bigg\{(\widehat{F}_{\g\rho}^B  {\a^B})\d_{\bar{\a}\bar{\b}}
\,+\,\widehat{F}_{\a\g}^{\bar{\xi}}f^{\bar{\xi}\rho\bar{\b}}
+{}\nn
\noalign{\vskip 0.2truecm}
&&\qquad\qquad +{}
\,\widehat{F}_{\rho\g}^{\bar{\xi}}f^{\bar{\xi}\a\bar{\b}}
\,+\,\widehat{F}_{\rho\a}^{\bar{\xi}}f^{\bar{\xi}\g\bar{\b}}\Bigg\}{\rho}^A
\,-\,\frac{2{\hbar}^{2}}{3}f^{\xi\g\rho}f^{\b\rho\a}
\widehat{F}_{\b\xi}^{\bar{\g}}{\g}^A\>.
\lab{tensorQCSA}
\ena
where we have used the following definitions:
\be
\widehat{F}_{\a\b}^{\bar{\g}}=\langle\, \widehat{F}_{\a\b}
\,,\,t^{\bar{\g}}\,\rangle,\quad
\widehat{F}_{\a\b}^{A} =\langle\,\widehat{F}_{\a\b} \,,\,t^{A}\,
\rangle,\quad 
\widehat{R}_{\a}^{\bar{\b}}=\langle\,\widehat{R}_{\a}
\,,\,t^{\bar{\b}}\,\rangle\,, \quad
\widehat{R}_{\a}^{A}=\langle\,\widehat{R}_{\a} \,,\,t^{A}\,\rangle.
\lab{proy}
\ee
Taking into account~\rf{tensorRCSA}--\rf{tensorQCSA} 
together with the constraints~\rf{const12} one can check that there 
is a particular solution for eq.~\rf{geni4} given by
\bea
&&{(\hbar k)^4}\widehat{F}_{\a\b}\,=\,
{(\hbar k)^4}\widehat{F}_{\a\b}^{(0)}(\vec{\l})\,
=-{\frac{{\d}_{\a\b}}{\,2\,
{\vec{\l}}{\cdot}{\vec{\a}}}}\> \vec\a \cdot\vec{t}.   
\lab{solF}\\
\noalign{\vskip 0.2truecm}
&&{{(\hbar k)}^3}\, \widehat{R}_{\a}\,=\,
{{(\hbar k)}^3}\Big\{\widehat{R}_{\a}^{(0)}(\vec{\l})
\,+\,
\frac{1}{k} \widehat{R}_{\a}^{(1)}(\vec{\l})\Big\},
\lab{solR}
\ena
where 
$\widehat{R}_{\a}^{(0)}(\vec{\l})$ is the classical
solution \rf{classRR} and 
\be
\widehat{R}_{\a}^{(1)}(\vec{\l})\,=\,
\frac{{\Psi_g^2}{h_g^{\lor}}\,-\,4{h_{g_0}^{\lor}}
\,+\,4\vec{\a}^2 }{4\>\vec{\l} \cdot \vec{\a}}
\> t^{\bar{\a}}.
\lab{solR1}
\ee
For this particular solution, eqs.~\rf{tensorR}--\rf{tensorQ} imply
\be
{(\hbar k)}^4 {R_{\a\b\g\rho}}\,
=\,{m}^3{R_{\a\b\g\rho}^{(0)}}(\vec{\l}),
\lab{soltensorR}
\ee
where $R_{\a\b\g\rho}^{(0)}(\vec{\l})$
is just the classical solution \rf{tensor11}, which is a consequence of
the absence of quantum corrections in eq.~\rf{tensorR}, 
\be
{(\hbar k)}^3 P_{\a\b\g}\,=\,-{m}^3\Big\{
 P_{\a\b\g}^{(0)}(\vec{\l})\,+
\,\frac{1}{k}P_{\a\b\g}^{(1)}(\vec{\l})\Big\},
\ee
where $P_{\a\b\g}^{(0)}(\vec{\l})$ is the classical solution given
by~\rf{tensor21} and the quantum correction $P_{\a\b\g}^{(1)}(\vec{\l})$ is 
\bea
&&-{m}^3 P_{\a\b\g}^{(1)}(\vec{\l})\,=\,
\frac{1}{6\,\vec{\l}\cdot\vec{\g}}\Bigg\{
\frac{{\Psi_g^2}h_g^{\lor}-4\,h_{g_0}^{\lor}}{4}\> \Big\{
 \frac{f^{\bar{\a}\b\bar{\g}}}
{\vec{\l}\cdot \vec{\a}}\,+\,\frac{f^{\bar{\b}\a\bar{\g}}}
{\vec{\l}\cdot \vec{\b}}\Big\}+{}
\nn
\noalign{\vskip 0.2truecm}
&&\qquad \qquad+{}
\frac{3f^{{\a}\b{\g}}}{2}\Big\{
\frac{\vec{\a}\cdot \vec{\g}}{\vec{\l}\cdot \vec{\a}}\,-\,
\frac{\vec{\b}\cdot \vec{\g}}
{\vec{\l}\cdot \vec{\b}}\Big\}
\Bigg\},
\ena
and finally
\be
{(\hbar k)}^2 Q_{\a\rho}\,=\,{m^3}\Big\{
Q_{\a\rho}^{(0)}(\vec{\l})\,+
\,\frac{1}{k}Q_{\a\rho}^{(1)}(\vec{\l})\,+
\,\frac{1}{k^2}Q_{\a\rho}^{(2)}(\vec{\l})\Big\},
\ee
where the first contribution is, again,
the classical one given by~\rf{tensor31} and the contributions 
$Q_{\a\rho}^{(1)}(\vec{\l})$ and $Q_{\a\rho}^{(2)}(\vec{\l})$ denote
respectively first and second order quantum corrections whose explicit
expressions are
\bea
&&{m^3}Q_{\a\rho}^{(1)}(\vec{\l})\,=\,
\frac{{\Psi_g}^2{h_g^\lor}-8{h_{g_0}^{\lor}}+
4{\vec{\a}}^2 }{8{(\vec{\l}\cdot\vec{\a})}^2}\> 
 \d_{\a\rho}\,+\,\frac{f^{\bar{\b}\g\bar{\rho}}
f^{\bar{\b}\g\bar{\a}}}{4(\vec{\l}\cdot\vec{\a})
(\vec{\l}\cdot\vec{\rho})}\> ,\\
\noalign{\vskip 0.4truecm}
&&{m^3}Q_{\a\rho}^{(2)}(\vec{\l})=
\frac{5(\vec{\g}\cdot\vec{\b})
\big\{ f^{\bar{\b}\g\bar{\rho}}f^{\bar{\g}\b\bar{\a}}\,+\,
 f^{\bar{\b}\g\bar{\a}}f^{\bar{\g}\b\bar{\rho}}\big\}}
{48(\vec{\l}\cdot\vec{\a})(\vec{\l}\cdot\vec{\rho})}\,+\,
{h_{g_0}^{\lor}}
\frac{{\Psi_g}^2{h_g^\lor}-4{h_{g_0}^{\lor}}+
4{\vec{\a}}^2 }{8{(\vec{\l}\cdot\vec{\a})}^2}\> \d_{\a\rho}+{}
\nn
\noalign{\vskip 0.2truecm}
&&\qquad\qquad
+{}\frac{f^{\bar{\b}\g\bar{\rho}}f^{\bar{\b}\g\bar{\a}}}
{4(\vec{\l}\cdot\vec{\a})(\vec{\l}\cdot\vec{\rho})}
\Big\{
\frac{{\Psi_g}^2{h_g^\lor}-4{h_{g_0}^{\lor}}}{4}\,-\,
\frac{\vec{\g}^2}{3}\,-\,
\frac{(\vec{\g}\cdot\vec{\b})(\vec{\l}\cdot\vec{\b})}
{2\> \vec{\l}\cdot\vec{\g}}\Big\}-{}
\nn
\noalign{\vskip 0.2truecm}
&&\qquad\qquad
-{}\frac{2 \vec{\rho}^4+
{(\vec{\b}\cdot\vec{\rho})}^2}{24{(\vec{\l}\cdot\vec{\rho})}^2}
\d_{\a\rho}  \>- \> 
\frac{f^{{\b}\g{\rho}}f^{{\b}\g{\a}}}
{8\> \vec{\l}\cdot\vec{\g}}\Big\{
\frac{\vec{\g}\cdot\vec{\rho}}{\vec{\l}\cdot\vec{\rho}}
+
\frac{\vec{\g}\cdot\vec{\a}}{\vec{\l}\cdot\vec{\a}}-
\frac{\vec{\g}\cdot\vec{\b}}{\vec{\l}\cdot\vec{\b}}
\Big\}.
\ena

The existence of this particular solution
for eq. \rf{geni4} means that,
at least, there is a spin-4 quantum conserved density or, equivalently, a
quantum conserved quantity of spin~3.  Moreover, as explained 
below~\rf{Lbar}, there will be also a conserved quantity of spin~$-3$.
Taking into account Parke's results~\cite{parke}, this allows one to
conclude that the Split models are quantum integrable and, hence, that they
should admit a factorizable $S$-matrix.

\sect{The spectrum of the Split models.}
\ \indent\label{soliton}

An important property of both the SSSG
and the HSG theories is that they admit soliton solutions~\cite{ntft}. 
Moreover, the semi-classical quantization of the solitons is expected to
help in deducing the form of the exact $S$-matrix. The complete analysis of
the soliton spectrum of the SSSG theories is beyond the scope of this paper
and will be presented in subsequent publications. However, in this section
we will discuss its main features and provide some explicit soliton
solutions.

We will restrict ourselves to the Split models, whose quantum integrability has
been explicitly established in the previous section. They are associated
with a type~I symmetric space $G/G_0$ of maximal rank, {\it i.e.\/}, ${\rm
rank\/}(G/G_0) = {\rm rank\/}(G)= r_g$, and two regular elements 
$\Lambda_\pm \in g_1$. This means that ${\rm Ker\/}({\ad\/}_{\Lambda_\pm})$
is a Cartan subalgebra of $g$ contained in $g_1$, and we will use the
realization of
$g_0$ and $g_1$ given in the Appendix. Actually, the simplest Split
model is the well known sine-Gordon model, which is recovered with the 
symmetric space $SU(2)/SO(2)$. In this case, $G_0$ is abelian and, hence,
the coupling constant does not have to be quantized; this is the reason why
this SSSG theory has not been mentioned in the
tables~\ref{table1}--\ref{table5}.

First of all, we have to obtain the vacuum manifold~${\cal M}_0$, which 
consists of the constant field configurations $h_0$ that minimise the
potential in~\rf{action}. This condition amounts to
\be
[\Lambda_+\>,\> h_0^\dagger \Lambda_- h_0]\> =\> 0\>, \quad {\rm and}\quad
\vec{\alpha}\bigl(\Lambda_+\bigr)\> \vec{\alpha}\bigl( h_0^\dagger \Lambda_-
h_0 \bigr) >0\>,
\lab{Minimum}
\ee
for all roots $\vec{\alpha}$ of $g$, which requires that $\Lambda_+$ and
$h_0^\dagger \Lambda_- h_0$ belong to the same Weyl chamber of the Cartan
subalgebra of $g$. Since they are regular, and taking into account that the
conjugation $h_0^\dagger \>\cdot\>  h_0$ permutes the Weyl
chambers~\cite{helgas}, we can assume that $\Lambda_+$ and $\Lambda_-$
already belong to the same Weyl chamber without any loss of
generality. Then~\rf{Minimum} implies
$h_0^\dagger \Lambda_- h_0= \Lambda_-$ and, therefore,
$h_0$ has to be of the form
\be
h_0\> =\> {\rm e\>}^{\pi \vec{\mu}\cdot \vec{t}}\>,
\lab{hcero1}
\ee
where $\vec{\mu}$ either vanishes or belongs to the co-root lattice of $G$,
which is the root lattice of the dual group $G^\vee$, $\Lambda_R(G^\vee)$.
Recall that $G^\vee$ is defined by requiring that its Lie algebra has roots
which are the duals of the roots of $g$ defined by $\vec{\alpha}^\vee =
2\>\vec{\alpha}/\vec{\alpha}^2$. However, any element of the form~\rf{hcero1}
satisfies $h_0^2 =1$, which implies that the vacuum manifold is given by
\be
{\cal M}_0\> =\> \bigl\{1\>, \>{\rm e\>}^{\pi \vec{\mu}\cdot \vec{t}}\mid
\vec{\mu}= \sum_{i=1}^{r_g} n_i \> \vec{\alpha}_i^\vee\>, n_i = 0,1\bigr\}\>,
\lab{VacMan}
\ee
where $\vec{\alpha}_1, \ldots, \vec{\alpha}_{r_g}$ are simple roots of $g$.
Therefore, ${\cal M}_0$ is an abelian discrete group isomorphic to the coset
$\Lambda_R(G^\vee)/2 \Lambda_R(G^\vee)$. Moreover, using the normalization in
the Appendix, one can check that
\be
{\rm e\>}^{\pi \vec{\alpha}^\vee \cdot \vec{t}}\> =\> \exp\left(+{2\pi\>
t^\alpha\over \sqrt{\vec{\alpha}^2}}\right)\> =\> \exp\left(-{2\pi\>
t^\alpha\over \sqrt{\vec{\alpha}^2}}\right)\>,
\lab{identity}
\ee 
for any root of $g$, which emphasises that ${\cal M}_0\subset G_0$.

\subsection{Fundamental particles.}
\ \indent

They  correspond
to the fluctuations of the field $h$ around a vacuum
configuration $h_0\in {\cal M}_0$. Let us take, $h=h_0 e^{\P}$, where $\P\in
g_0$. The linearized eqs.~\rf{equation12} are
\be
P(\P)\,=\,0 \quad {\rm and} \quad \partial_{\m}\partial^{\m}\P
\,=\,-4{m^2}[\,\Lambda_{+} ,[\Lambda_{-}\>, \> \P] ],
\lab{linear}
\ee
which show that the fundamental particles are associated with the non-vanishing
eigenvalues of the mass-matrix $-4{m^2}[\Lambda_{+},[ \Lambda_{-} ,\P]]$ on
$g_0$. They correspond to the field configurations of the form
$\P=\phi t^{\a}$, where $\vec\a$ is an arbitrary positive root of
$g$, whose mass is
\be
m_{\a}\> =\> 2m\sqrt{\vec{\alpha}\bigl(\Lambda_+\bigr)\> \vec{\alpha}\bigl( 
\Lambda_- \bigr)},
\lab{massesf}
\ee
which is real because of~\rf{Minimum}. Therefore, for each positive root
$\vec{\alpha}$ of $g$, there is a fundamental particle described by a real
field $\phi=\phi(x,t)$ whose mass is given by~\rf{massesf}. It is worth
noticing that~\rf{massesf} is the mass formula giving the spectrum of
fundamental particles of the HSG theory associated with~$G$; however, in that
case the particles are described by complex fields.

A very peculiar property of this spectrum that is shared with the HSG 
and all the other SSSG theories is that the mass formula~\rf{massesf} satisfies
the following kind of inequalities
\be
m_{\a+\b} \geq  m_{\a} \,+\, m_{\b}\> ,
\lab{funineq}
\ee
which suggests that some of the fundamental particles might be unstable.
For the HSG theories, this has been checked using perturbation
theory~\cite{hsg3} and, consequently, their exact $S$-matrix exhibits resonance
poles associated with the unstable particles~\cite{hsg3,TBA}. 
For the Split model related to $SU(3)/SO(3)$ it has also been checked that some
of the fundamental particles actually decay~\cite{brazhnikov};
however, more work is needed in order to establish this for the generality of
the SSSG models and construct their exact $S$-matrix.

\subsection{Soliton solutions.}
\ \indent

In order for a solution to eqs.~\rf{equation12} to have
finite energy the field $h$ must tend to limits in ${\cal M}_0$ as $x\ra
\pm\infty$. So,
\be
h(+\infty, t)\> h^\dagger(-\infty, t) \in {\cal M}_0\>,
\lab{boundary}
\ee
and its value is conserved as the system evolves in time. This means that, 
at fixed $t$, a solution $h=h(t,x)$ with finite energy is a path on the $G_0$
manifold connecting two elements in ${\cal M}_0$ and, since
$G_0$ is not simply connected in general, there could be
different solutions sharing the same value of $h(+\infty, t)
h^\dagger(-\infty, t)$. Therefore, each solution will be characterized by two
topological `quantum numbers'. The first will be the value of~\rf{boundary},
which is an element in ${\cal M}_0$ or, equivalently, in
the discrete group $\Lambda_R(G^\vee) /2
\Lambda_R(G^\vee)$, and the second will be an element in $\pi_1(G_0)$, the
fundamental group of $G_0$, which can be found in table~\ref{table6}. In other
words, the solitons of the Split models will be topological, like the solitons of
the sine-Gordon equation or of the affine Toda equations with imaginary 
coupling constant. This is in contrast with the solitons of the
HSG theories, which only carry Noether charges. On their side, the solitons of
the Split models do not carry any Noether charge because $g_0^0=\{0\}$
in~\rf{lambcond} ($p=0$). Nevertheless, for a generic SSSG theory
$p\not=0$ and the solitons will carry both topological and
$U(1)^p$ Noether charges, which make them similar to the dyons in
four-dimensional non-abelian gauge theories~\cite{dyon}

\begin{table}[!ht]
\begin{center}
\begin{tabular}{|c||c|c|c|c|c|c|c|} \hline
& & &
&$Sp(4)$\\ 
${G_0}$& $SO(n)$&${SO(n) \times SO(n+1)}$&$U(n)$&$SU(8)$\\ 
&$SO(16)$&${SO(n) \times SO(n)}$&&${Sp(3) \times SU(2)}$\\ 
&&&&${SU(2)  \times SU(2)}$\\ \hline
&&&&\\ 
${\pi_1}{({G_0})}$&$\ZZ_2$&$\ZZ_2 \oplus \ZZ_2$&
$\ZZ$&$0$\\ 
&&&&\\ \hline
\end{tabular}
\end{center}
\caption{Fundamental groups ${\pi_1}{({G_0})}$ corresponding to the Split
models.}
\label{table6}
\end{table}

Taking into account that the sine-Gordon theory is the Split model
corresponding to $SU(2)/SO(2)$, a number of explicit soliton solutions for
the Split models can be obtained by embedding the sine-Gordon solitons in the
regular $SU(2)$ subgroups of $G$. This method is widely used in the context
of Yang-Mills theories based on arbitrary Lie groups to construct monopole or
instanton solutions by embeddings of the $SU(2)$ spherically symmetric
't-Hooft-Polyakov monopole~\cite{Monopole} or the self-dual $SU(2)$
Belavin-Polyakov-Schwartz-Tyupkin instanton~\cite{Instanton}. It has also been
used to construct the soliton solutions of the affine Toda theories with
imaginary coupling constant~\cite{TodaJap} and, more recently, to construct the
soliton solutions of the HSG theories starting with the Complex sine-Gordon
solitons~\cite{hsg2}. For each positive root $\vec{\alpha}$ of~$G$, let us
consider the field configuration
$h=\exp(\phi t^\alpha/\sqrt{\vec{\alpha}^2})$, which trivially satisfies the
constraints in~\rf{equation12}. According to~\rf{action}, its Lagrangian
density is
\be
{\cl} \,=\,
\frac{1}{4 \pi {\b^2} \vec{\a}^2} 
\Bigl(\frac{1}{2} \partial_{\mu}\p \> \partial^{\mu}\p 
\,+\, m_\a^2\bigl(\cos \p\, -\,1 \big) \Bigr),
\lab{lagrangian}
\ee 
which is just the Lagrangian density of the sine-Gordon model.

This way, for each positive root $\vec{\alpha}$ of $G$, each soliton solution of
the sine-Gordon equation provides a soliton solution for the Split model.
Namely, the usual soliton and anti-soliton solutions allow one to construct
two {\it a priori\/} different soliton solutions with mass
\be
M_{s,\bar{s}}({\vec\a})=\frac{2}{\vec{\a}^2}\,\frac{1}{\pi {\b^2}}\,  
m_{\a},
\lab{masses}
\ee
while the masses of the solitons associated with the breathers of
the sine-Gordon equation are 
\be 
M_n(\vec\a)\,=\,2M_s({\vec\a})\> \sin \Big({\vec{\a}^2 \over2} 
\frac{\pi \beta^2}{2}
\, n\Big)\,\leq \,2M_s({\vec\a}), 
\lab{mbreathers}
\ee 
and there is a different soliton for each integer $n$ such that $n<
2/(\beta^2 \vec{\a}^2)$. As usual, in the weak coupling $\beta^2 \ra 0$ limit we
obtain
\be
M_n(\vec\a)\,\simeq \,  n m_{\a}-
{\cal O} \Big(\beta^2\Big) 
\lab{semic}
\ee 
and, hence, the fundamental particle associated with $\vec\a$ becomes identified
with the lightest breather with mass $M_1(\vec{\alpha})$. It is important to
notice that the mass formulae~\rf{masses} and~\rf{mbreathers} satisfy
inequalities similar to~\rf{funineq}, which again suggests that some of
these solitons might be unstable. Actually, some examples of unstable solitons 
are already known in the Split model related to $SU(3)/SO(3)$ ~\cite{brazhnikov}.

The soliton ($s$) and antisoliton ($\bar{s}$)
solutions of the sine-Gordon model satisfy
\be
\p_{s, \bar{s}}(+\infty,t)\,-\,
\p_{s, \bar{s}}(-\infty,t)\,=\,\pm 2 \pi,
\lab{asimp1}
\ee
which implies the following asymptotic behaviour for the fields
$h_s=\exp(\phi_s t^\alpha/\sqrt{\vec{\alpha}^2})$ and
$h_{\bar{s}}=\exp(\phi_{\bar{s}} t^\alpha/\sqrt{\vec{\alpha}^2})$:
\be
h_s(+\infty,t)\> h_s^\dagger(-\infty,t) \> =\> {\rm e\>}^{+\frac{2 \pi t^{\a}}
{\sqrt{\vec{\a}^2}}} , \qquad
h_{\bar s}(+\infty,t)\> h_{\bar s}^\dagger(-\infty,t) \> =\>  {\rm
e\>}^{-\frac{2 \pi t^{\a}} {\sqrt{\vec{\a}^2}}} \>,
\lab{asimpt2}
\ee
both in ${\cal M}_0$. Therefore, taking into account~\rf{identity}, their
asymptotic behaviour is the same and it will not be possible to distinguish
these two soliton configurations unless $G_0$ is not simply connected
and $h_s$ and $h_{\bar{s}}$ are associated with different elements in
$\pi_1(G_0)$.

\sect{Conclusions.}
\ \indent\label{conclusions}

In this paper we have studied some of the quantum properties of the massive
SSSG theories constructed in~\cite{ntft}. First of all, we
have identified the perturbed conformal field theories corresponding to these
theories when the symmetric space $G/G_0$ is of type~I. This amounts to find
which are the unperturbed CFT and the perturbing operator specified by the
potential term in the classical action. Since the type~I symmetric spaces are
irreducible, the perturbation is given by a single spinless
primary field whose conformal dimension has been calculated and
can be found in tables~\ref{table1} and~\ref{table2}. Actually, our
calculation only depends on the algebraic structure of
the symmetric space and, therefore, it provides the conformal dimension of the
perturbations for the general class of SSSG theories constructed
in~\cite{sssg3}. They are obtained from ours by substituting $\Lambda_+$ and
$\Lambda_-$ for two arbitrary elements $T$ and
$\bar{T}$ in $g_1$, and $g_0^0$ for ${\go h}$, their simultaneous centralizer in
$g_0$. The resulting SSSG theories are perturbations of the coset CFT related to
$G/H$, where $H$ is the Lie group corresponding to $\go h$. However, as shown 
in~\cite{ntft}, these theories will not exhibit a mass gap unless $H$ is either
trivial or abelian.

Among others, the resulting class of perturbed CFT's
include massive perturbations of WZNW models, which are related to
symmetric spaces of maximal rank; we have named these theories `Split
models'. In addition, there are new massive perturbations of parafermion
theories different to those provided by the HSG theories~\cite{hsg1}. In
particular, this class of theories includes the perturbations of the
simplest $\ZZ_k$ parafermions by their first and second thermal
operators~\cite{paras}, which are related to
$G/G_0 = Sp(2)/U(2)$~\cite{ls} and $SU(3)/SO(3)$~\cite{brazhnikov},
respectively. 

Our second task was to investigate the quantum integrability of the SSSG
theories. In view of the large variety of different types of perturbed CFT's
corresponding to the SSSG theories, we have restricted ourselves to
give a detailed proof of the quantum integrability only for
the Split models. Classically, they exhibit ${\rm rank\/}(G)$ conserved
quantities for each odd spin $\pm1, \pm3,\ldots$, and we have checked that there
are at least two quantities of spin $+3$ and
$-3$ that remain conserved in the quantum theory after an appropriate
renormalization. This implies, via the usual folklore, the factorization of
their scattering matrices~\cite{parke} and, hence, their quantum integrability.
This result, together with the  integrability of the HSG theories~\cite{hsg1}
whose simplest higher spin conserved quantities have spin $\pm2$, lead us to
conjecture that all the massive SSSG theories will be quantum integrable.

The quantum integrability of the SSSG theories implies that they should admit a
factorizable $S$-matrix and the next stage of analysis consists in establishing
its form. As a first step towards this aim, we have illustrated the general
properties of the spectrum of the SSSG theories by discussing the spectrum of
fundamental particles and solitons of the Split models. Its main features,
which are expected to be shared with all the other SSSG theories, are, first,
that the fundamental particles become identified with some of the solitons in the
semiclassical limit, like in the sine-Gordon and complex sine-Gordon theories.
This makes us expect that the spectrum will be entirely solitonic in the general
case. Second, since there are different vacua, these solitons are topological,
in contrast with the solitons of the HSG theories which are not. Moreover, in
the general case, they are expected to carry conserved Noether charges as
well. Third, some of these solitons are expected to correspond to unstable
particles in the quantum theory. Therefore, like in the HSG theories, 
only the stable solitons should correspond to asymptotic states while the
unstable ones will produce resonance poles in the $S$-matrix~\cite{hsg3,TBA}.
Finally, both the fundamental particles and the solitons are labelled by the
roots of the Lie algebra of $G$, which is reminiscent of the spectrum of the
HSG theories. 

Taking into account the form of the exact $S$-matrices of the HSG
theories~\cite{hsg3,TBA}, the last feature suggests that the
$S$-matrices of the  SSSG theories could be somehow related to the 
`colour valued' scattering matrices constructed in~\cite{Andreas}. These
$S$-matrices are related to pairs $\{\tilde{g}|g\}$ of simply laced Lie
algebras, where $\tilde{g}$ governs the mass spectrum and the fusing rules,
while
$g$ provides the `colour' quantum numbers. Using this construction, the
$S$-matrix of the HSG theory corresponding to the Lie group $G$ at level
$k$ is related to the
pair $\{A_{k-1}| g\}$, where $g$ is the Lie algebra of~$G$. Then,
it is worthwhile to notice that the conjectured~\cite{Andreas} central
charge of the ultraviolet CFT corresponding to the colour valued $S$-matrix
specified by the pair
$\{D_k|g\}$ coincides with the central charge of the unperturbed CFT
of the $G/G_0$ Split model ($p=0$) at level~$k$, or level~$2k$ if
$G=SU(m)$, where $G/G_0$ is one of the symmetric spaces listed
in~\rf{SplitSS} with simply laced $G$. However, additional work is needed
in order to go beyond this numerical coincidence.

\vspace{0.75 truecm}

\noindent\centerline{\large\bf Acknowledgments} 

\vspace{0.25truecm}
We would like to thank J.~S\'anchez Guill\'en, A.~Fring and C.~Korff for
their valuable comments. This research is supported partially by CICYT
(AEN99-0589), DGICYT (PB96-0960), and the EC Commission via a TMR Grant
(FMRX-CT96-0012).

\vspace{1 cm}

\appendix
\sect{Classical conserved densities of the Split models.}
\label{App}                               

\indent \ \

In this appendix we give the explicit expressions for the classical 
conserved densities with scale dimension~2 and~4 corresponding to the Split
models, which have been obtained by solving eqs.~\rf{equation12}.

We will use the standard explicit realization of
the basis $t^a$ of $g$ in terms of a Cartan basis of its
complexification
\bea
&&t^A\,=\,i H^A\>, \qquad A\> =\> 1, \ldots, {\rm rank\/}(g)\>,\nn
\noalign{0.2truecm} 
&&{t}^{\a}\,=\,\sqrt{\frac{1}
{2\langle\,{E}_{\a},{E}_{-\a}\,\rangle}}
\> ({E}_{\a}-{E}_{-\a})\>, \quad
{t}^{\bar{\a}}\,=\, i\> \sqrt{\frac{1} {2\langle\,{E}_{\a},{E}_{-\a}\,
\rangle}}
\> ({E}_{\a}+{E}_{-\a}),
\lab{basis}
\ena
where the $H^A$'s provide an orthonormal basis for
the Cartan subalgebra with respect to the invariant bilinear form
$\langle\>, \> \rangle$, and the step operators
$E_\alpha$ are normalized such that
\be
[\,H^A,E_{\a}\,]\,=\,\vec{\alpha}(H^A)\> E_{\a} \,=\,\alpha^A\> E_{\a}\>,
\qquad [\,E_{\a},E_{-\a}\,]\,=\,\langle\,E_{\a},E_{-\a}\rangle \,\alpha^A
H^A,
\lab{chevalley}
\ee
with $\vec{\alpha}$ a positive root of~$g$.

The generators~\rf{basis} are normalized in such a way
that $\langle \,t^a, t^b\, \rangle= -\d^{ab}$.
Moreover, the corresponding structure functions are totally
antisymmetric and satisfy:
\bea
&&f^{\bar{\a}\bar{\b}\bar{\g}}=f^{\a\b\bar{\g}}=
f^{\a\bar{\a}\b}=f^{\a\bar{\a}\bar{\b}}=0,\,\,
\,\,\forall \,\,\vec\a,\vec\b,\vec\g>0,
\nn
\noalign{\vskip 0.2truecm}
&&f^{ABC}=f^{AB\a}=f^{AB\bar{\a}}=0,\,\,\,\,\forall 
\,\,\vec\a>0\,\,\,\,  {\rm and } \,\,\,\,
\forall\,\, \,\,A,B=1,\cdots , {\rm rank\/}(g),
\nn
\noalign{\vskip 0.2truecm}
&&f^{\a\bar{\a}A}=\a^{A},\,\,\,\,\forall \,\,\vec\a>0 \,\,\,\,
{\rm  and}\,\,\,\,\forall\,\, \,\,
 A=1,\cdots, {\rm rank\/}(g).
\lab{struc123} 
\ena

Recall that the rank of a symmetric space $G/G_0$ is the dimension of any maximal
abelian subspace contained in $g_1$. Therefore, if ${\rm rank\/}(G/G_0)= {\rm
rank\/}(G)$, one can take $t^A \in g_1$ for all $A$ and, hence, the
subset of generators ${t}^{{\a}}$ provide a basis for the subalgebra $g_0$,
while the ${t}^{\bar{\a}}$'s, together with the $t^A$'s, generate
$g_1$. We will use this realization of $g_0$ and $g_1$ in our calculations for
the Split models. Therefore, 
\be
\L_{-}\> =\> \l^A \> t^A \>=\> \vec{\l}\cdot \vec{t}\>, 
\lab{Lambase}
\ee
and the potential
$q=q(x)$ in the Lax operator~\rf{lax1} is of the form $q=q^\a t^\a$. 

Taking into account all this, the expressions
for the conserved densities of scale-dimension~2 are
\be
\m^A {\cal I}_2^{(0)A}\,=\,\vec{\m} \cdot\vec{{\cal
I}}_2^{(0)}\,=\,D_{\a\b}q^{\a}q^{\b},\quad {\rm with}\quad
D_{\a\b}\,=\,{1\over 2m}\> \frac{\vec{\m}\cdot\vec{\a}}
{\vec{\lambda}\cdot\vec{\a}},
\lab{spin2class}
\ee
where $\vec\m$ is an arbitrary ${\rm rank\/}(g)$-component vector  which
allows one to write the ${\rm rank\/}(g)$ conserved quantities ${\cal
I}_2^{(0)A}$  in a compact way. In particular, for $\vec\m=\vec\l$,
\rf{spin2class} reduces to 
\be
m{\lambda^A} {\cal I}_{2}^{(0)A}\,=\,
\frac{1}{2} q^{\a}q^{\a}\,=\,2\pi \b^2 T_{--},
\lab{spin22class}
\ee
which is one of the components of the classical stress-energy tensor.

For scale dimension~4, the ${\rm rank\/}(g)$-conserved densities can be
written as:
\be
\vec{\m}\cdot\vec{\cal I}_{4}^{(0)}\,=\,R_{\a\b\g\rho}^{(0)}(\vec{\m})
q^{\a} q^{\b} q^{\g} q^{\rho}
\,+\,P_{\a\b\g}^{(0)}(\vec{\m}) q^{\a} q^{\b} \partial_{-} q^{\g}\,+\,
Q_{\a\b}^{(0)}(\vec{\m})q^{\a}\partial_{-}^{2} q^{\b},
\lab{spin4class}
\ee
where
\bea
&&R_{\a\b\g\rho}^{(0)}(\vec{\m})=\bigg( -{\frac{({\vec 
{\a}}\cdot{\vec{\g}}) ({\vec{\g}}\cdot{\vec{\m}})} {{8 { m ^{ 
3}}}{({\vec{\l}}\cdot{\vec{\a}})({\vec{\l}}\cdot{\vec{\g}})^{2}}}}
 \, {{\d}_{\a\b}} {{\d}_{\g\rho }}   
-{\frac{f^{{\bar{\b}} {\g} {\bar{\xi}}} f ^{\xi\rho\a} \, 
({\vec{\a}}\cdot{\vec{\m}})} {6{ m ^{ 3}} ({\vec 
{\l}}\cdot{\vec{\b}})({\vec{\l}}\cdot{\vec{\xi}})({\vec{\l}}\cdot{\vec{\a}})}}
+{} 
\nn                                                                                                                             
\noalign{\vskip 0.2truecm}
&& \qquad{}+  {\frac{f^{\bar{\rho}\a\bar{\xi}} f ^{{\bar{\xi}}{\bar{\g}}\b}} 
{{24 m^{ 3}}({\vec{\l}}\cdot{\vec{\g}}) ({\vec {\l}}\cdot{\vec{\rho}}) } 
\bigg\{-{\frac {{\vec{\b}}\cdot{\vec{\m}}} {{\vec {\l}}\cdot{\vec{\b}}}} + 
{\frac{\vec{\xi}\cdot\vec{\m}} {\vec{\l}\cdot\vec{\xi}}}\bigg\}}
\bigg)_{{\rm sym\/}(\a\b\g\rho)},
\lab{tensor1} \\
\noalign{\vskip 0.4truecm}
&&P_{\alpha\beta\gamma}^{(0)}(\vec{\m})=\biggl({\frac{f ^{\g\a\b}} {3{ m ^{ 
3}} ({\vec{\l}}\cdot{\vec{\g}})^{2}}}\,{\frac{{\vec{\b}}\cdot{\vec{\m}}}
{{\vec  {\l}}\cdot{\vec{\b}}}} + \frac{f ^{\bar{\a}\g\bar{\b}}} {4{ m ^{ 
3}}({\vec{\l}}\cdot{\vec{\a}})({\vec{\l}}\cdot{\vec{\b}})}\,\frac{
{\vec{\b}}\cdot{\vec{\m}}} {{\vec{\l}}\cdot{\vec{\b}}}-{}\nn
\noalign{\vskip 0.2truecm}
&& \qquad{}- {\frac{f ^{\bar{\b}\a\bar{\g}}} {4{ m ^{ 
3}}({\vec{\l}}\cdot{\vec{\g}})({\vec{\l}}\cdot{\vec{\b}})}}
\bigg\{{\frac{{\vec{\g}}\cdot{\vec{\m}}} {\,3\,{\vec{\l}}\cdot{\vec{\g}}}}
+{\frac{{\vec{\b}}\cdot{\vec{\m}}} 
{{\vec{\l}}\cdot{\vec{\b}}}}\bigg\}\biggr)_{{\rm sym\/}(\a\b)},
\lab{tensor2}\\
\noalign{\vskip 0.4truecm}
&&Q_{\alpha\beta}^{(0)}(\vec{\m})=-{\frac{{\vec{\a}}\cdot{\vec{\m}}}
{\,2\, { m ^{ 3}}{({\vec{\l}}\cdot{\vec{\a}})^{ 3}}}}\, {\d}_{\alpha\beta}.
\lab{tensor3}
\ena
In these equations, ${\rm sym\/}(\a\b\g\rho)$ and  ${\rm
sym\/}(\a\b)$ means that the corresponding expressions have to be considered
completely symmetrized in $(\a\b\g\rho)$ or $(\a\b)$, respectively. In
particular, for $\vec\m=\vec\l$ these expressions simplify to
\bea
&&{R_{\a\b\g\rho}^{(0)}}(\vec{\l})= - {\frac{1}{{24m ^{ 
3}}\> {\vec{\l}}\cdot{\vec{\a}}}}\left\{{\frac{{\vec 
{\a}}\cdot{\vec{\g}}}{{\vec{\l}}\cdot{\vec{\g}}}}  \,\left\{ {{\d}_{\a\b}} 
{{\d}_{\g\rho }}\,+\,{{\d}_{\a\rho}} {{\d}_{\g\b}} \right\} 
\,+\,{\frac{{\vec 
{\a}}\cdot{\vec{\b}}}{{\vec{\l}}\cdot{\vec{\b}}}}{{\d}_{\a\g}} 
{{\d}_{\b\rho}}\, \right\},
\lab{tensor11}\\
\noalign{\vskip 0.2truecm}
&&{P_{\alpha\beta\gamma}^{(0)}}(\vec{\l})= {\frac{-1}{{6 m 
^{3}}\>{\vec{\l}}\cdot{\vec{\g}}}}\left\{{\frac{f ^{\bar{\b}\a\bar{\g}}} 
{{\vec{\l}}\cdot{\vec{\b}}}}+ {\frac{f ^{\bar{\a}\b\bar{\g}}}
 {{\vec{\l}}\cdot{\vec{\a}}}}\right\},
\lab{tensor21}\\
\noalign{\vskip 0.2truecm}
&&{Q_{\alpha\beta}^{(0)}}(\vec{\l})=
-{\frac{{\d}_{\alpha\beta}}{\,2\, 
{ m ^{ 3}}{({\vec{\l}}\cdot{\vec{\a}})^{ 2}}}}.
\lab{tensor31}
\ena

\ \indent
\vspace{1.5cm}

\begin{table}[h]
\begin{center}
\begin{tabular}{|c|c|c|}\hline
&&\\
$G/{G_0}$&rank($G/{G_0}$)&$\Delta_{\Phi}$\\
&&\\
\hline\hline
&&\\
$SU(3)/SO(3)$&2& $\frac{6}{k+2}$\\
&&\\
\hline
&&\\
$SU(4)/SO(4)$&3&$\frac{4}{k+2}$\\
&&\\
\hline
&&\\
$SU(n)/{SO(n)}$, \,\, $(n \geq 5)$&$n-1$&$\frac{n}{2(k+n-2)}$\\
&&\\
\hline
&&\\
$SU(2n)/Sp(n)$, \,\, $(n \geq 1)$&$ n-1$&$\frac{n}{k+n+1}$\\

&&\\
\hline
&&\\
${SO(n+2)}/{SO(n) \times U(1)}$, \,\, $(n \geq 5)$&
$2$&
$\frac{n-1}{2(k+n-2)}+\frac{1}{2k}$\\
&&\\
\hline
&&\\
${SO(n+3)}/{SO(n) \times SO(3)}$, \,\, $(n \geq 4)$&
3&
$\frac{n-1}{2(k+n-2)}+\frac{1}{k+2}$\\
&&\\
\hline
&&\\
${SO(n+m)}/{SO(n) \times SO(m)}$, \,\, $(n, m \geq 4)$&
min$(n,m)$&
$\frac{n-1}{2(k+n-2)}+\frac{m-1}{2(k+m-2)}$\\
&&\\
\hline
&&\\
$Sp(n+m)/{Sp(n) \times Sp(m)}$, \,\, $(n,m  \geq 1)$&
min$(n,m)$&
$\frac{1+2n}{4(k+n+1)}+\frac{1+2m}{4(k+m+1)}$\\
&&\\
\hline
&&\\
${SU(n+m)}/{SU(n)\times SU(m)\times U(1)}$,& min$(n,m)$&
$\frac{(n-1)(n+1)}{2n(k+n)}+
\frac{(m-1)(m+1)}{2m(k+m)}+\frac{n+m}{2nmk}
$\\
$(n, m \geq 2)$& &\\
&&\\
\hline
&&\\
${Sp(n)}/{SU(n) \times U(1)}$, \,\, $(n  \geq 2)$&n&
$\frac{(n-1)(n+2)}{n(k+n)}+\frac{2}{nk}$\\
&&\\
\hline
&&\\
${SO(2n)} /{SU(n) \times U(1)}$, \,\, $(n  \geq 3)$&$[n/2]$&
$\frac{n^2-n-4}{2n(k+n-1)} + \frac{2}{nk}$\\
&&\\
\hline
\end{tabular}
\end{center}
\caption{Conformal dimensions of the perturbations corresponding to the type~I 
SSSG-models associated to the classical Lie groups $G$.}
\label{table1}
\end{table}

\ \indent
\vspace{1.7cm}

\begin{table}[h]
\begin{center}
\begin{tabular}{|c|c|c|}\hline
&&\\
$G/{G_0}$&rank($G/{G_0}$)&$\Delta_{\Phi}$\\
&&\\
\hline\hline
&&\\
${E_6}/{Sp(4)}$& 6&
 $\frac{6}{k+5}$\\
&&\\
\hline
&&\\
${E_6}/{F_4}$&2&$\frac{6}{k+9}$\\
&&\\
\hline
&&\\
${E_7}/{SU(8)}$&7&$\frac{9}{k+8}$\\
&&\\
\hline
&&\\
${E_8}/{SO(16)}$&8&$\frac{15}{k+14}$\\
&&\\
\hline
&&\\
${F_4}/{SO(9)}$&1&$\frac{9}{2(k+7)}$\\
&&\\
\hline
&&\\
${E_6}/{SU(6) \times SU(2)}$&4&
$\frac{21}{4(k+6)}+\frac{3}{4(k+2)}$
\\
&&\\
\hline
&&\\
${E_7}/{SO(12)\times SU(2)}$&4&$
\frac{33}{4(k+10)}+ \frac{3}{4(k+2)}$\\
&&\\
\hline
&&\\
${E_8}/{E_7 \times SU(2)}$&4&$\frac{57}{4(k+18)}+ \frac{3}{4(k+2)}$\\
&&\\
\hline
&&\\
${F_4}/{Sp(3)\times SU(2)}$&4&
$\frac{15}{4(k+4)}+\frac{3}{4(k+2)}$
\\
&&\\
\hline
&&\\
${G_2}/{SU(2)\times SU(2)}$&2
&$\frac{15}{4(3k+2)}+\frac{3}{4(k+2)}$
\\
&&\\
\hline
&&\\
${E_6}/{SO(10) \times U(1)}$& 2& $ \frac{767}{128(k+8)}+
 \frac{1}{128k}$\\
&&\\
\hline
&&\\
${E_7}/{E_6 \times U(1)}$& 3&  $ \frac{527}{72(k+12)}+
\frac{1}{72k}$\\
&&\\
\hline
\end{tabular}
\end{center}
\caption{Conformal dimensions of the perturbations corresponding to the 
type~I SSSG-models associated to the exceptional Lie groups $G$.}
\label{table2}
\end{table}

\ \indent

\vspace{2cm} 

\begin{table}[h]
\begin{center}
\begin{tabular}{|c|c|c|c|}\hline
&&&\\
$G/{G_0}$ & Type &$\vec{s}$&
$\tilde{C}_2 (g^{(i)})=C_2 (g^{(i)}) /{\vec\Psi_{g^{(i)}}^2}$\\
&&&\\
\hline\hline
&&&\\
$SU(3)/SO(3)$&A2&$(0,1)$&6\\
&&&\\
\hline
&&&\\
$SU(4)/SO(4)$&A2&$(0,1,0)$&$\tilde{C}_2 (g^{(1)})=\tilde{C}_2 (g^{(2)})=2$\\
&&&\\
\hline
&&&\\
${SU(n)}/{SO(n)}$,\,\,$(n\geq 5)$&A2&
$(0,\ldots,0,1)$& $n$\\
&&&\\
\hline
&&&\\
${SU(2n)}/{Sp(n)}$,\,\,$(n\geq 1)$&A2&
$(1,\ldots,0)$& $n$\\
&&&\\
\hline
&&&\\
${SO(n+3)}/{SO(n) \times SO(3)}$&
A1 &$(0,\ldots,1,0)$&
$\tilde{C}_2 (D_p)=p-1/2$\\
$n=2p,\,\,(p \geq 2)$  && 
&$\tilde{C}_2 (B_1) =1$\\
&&&\\
\hline
&&&\\
${SO(n+3)}/{SO(n) \times SO(3)}$&
A2 &$(0,\ldots,1,0)$&
$\tilde{C}_2 (B_p)=p$\\
$n=2p+1,\,\,(p \geq 2)$  && 
&$\tilde{C}_2 (B_1) =1$\\
&&&\\
\hline
&&&\\
${SO(n+m)}/{SO(n) \times SO(m)}$&
A1 &
$(0,\ldots,1,\ldots,0)$,&
$\tilde{C}_2 (D_p)=p-1/2$\\
$n=2p, m=2q, \,\,(p,q \geq 2)$  && $s_{p}=1$
&$\tilde{C}_2 (D_q) =q-1/2$\\
&&&\\
\hline
&&&\\
${SO(n+m)}/{SO(n) \times SO(m)}$&
A1 &
$(0,\ldots,1,\ldots,0)$, &
$\tilde{C}_2 (D_p)=p-1/2$\\
$n=2p, m=2q+1,\,\,(p, q \geq 2) $
&&$s_{p}=1$
&$\tilde{C}_2 (B_q)=q$\\
&&&\\
\hline
&&&\\
${SO(n+m)}/{SO(n) \times SO(m)}$&
A2 &
$(0,\ldots,1,\ldots,0)$, &
$\tilde{C}_2 (B_p)=p$\\ 
$n=2p+1, m=2q+1,\,\,(p, q \geq 2)$ &&$s_{p}=1$
&$\tilde{C}_2 (B_q) =q$\\
&&&\\
\hline
&&&\\
${Sp(n+m)}/{Sp(n) \times Sp(m)}$,\,\, $(n, m \geq 1)$& A1&
$(0,\ldots,1,\ldots,0),$&
$\tilde{C}_2 (C_n)=(2n+1)/4$ \\
&&$s_n=1$&$\tilde{C}_2 (C_m)= (2m+1)/4$\\ 
&&&\\
\hline
\end{tabular}
\end{center}
\caption{Type, $\vec{s}$, and $C_2 (g^{(i)}) /{\vec\Psi_{g^{(i)}}^2}$
corresponding to the SSSG-models of types~A1
and~A2 associated to $G$ classical.}
\label{table3}
\end{table}

\ \indent
\vspace{3cm}

\begin{table}[h]
\begin{center}
\begin{tabular}{|c|c|c|c|}\hline
&&&\\
$G/{G_0}$ & Type &$\vec{s}$&
$\tilde{C}_2 (g^{(i)}) =C_2 (g^{(i)})/{\vec\Psi_{g^{(i)}}^2}$\\
&&&\\
\hline\hline
&&&\\
${SO(n+2)}/{SO(n) \times U(1)}$&B&
$(1,1,0\ldots,0)$&
$\tilde{C}_2 (D_p)=p-1/2$\\
$n=2p, \,\,(p \geq 3)$ &&&
$\vec{\Lambda} \cdot \vec{u}={\vec\Psi_{D_p}^2}/2$ \\
&&&\\
\hline
&&&\\
${SO(n+2)}/{SO(n) \times U(1)}$&B&
$(1,1,0\ldots,0)$&
$\tilde{C}_2 (B_p)=p$\\
$n=2p+1,\,\, (p \geq 2)$
&&&$\vec{\Lambda} \cdot \vec{u}={\vec\Psi_{B_p}^2}/2$ \\
&&&\\
\hline
&&&\\
${SU(n+m)}/{SU(n)\times SU(m) \times U(1)}$&&$(1,0,\ldots,1,\ldots,0)$&
$\tilde{C}_2(A_{n-1})=
\frac{(n-1)(n+1)}{2n}$\\
$(n,m \geq 2)$&B&$s_0=s_n=1$& 
$\tilde{C}_2(A_{m-1})=
\frac{(m-1)(m+1)}{2m}$\\
&&&
$\vec\Lambda\cdot\vec{u}=
\frac{n+m}{2nm}\vec\Psi_{A_{n-1}}^2 $\\
&&&\\
\hline
&&&\\
${Sp(n)}/{SU(n) \times U(1)}$&B&
$(1,0,\ldots,0,1)$&$
\tilde{C}_2(A_{n-1})=\frac{(n-1)(n+2)}{n}$\\
$(n \geq 2)$ &&& $\vec\Lambda\cdot\vec{u}=\frac{2}{n} 
\vec\Psi_{A_{n-1}}^2$\\
&&&\\
\hline
&&&\\
${SO(2n)}/{SU(n)\times U(1)}$&B&
$(1,0,\ldots,0,1)$&$
\tilde{C}_2(A_{n-1})=\frac{(n^2-n-4)}{2n}$\\
$(n \geq 3)$&&&
$\vec\Lambda\cdot\vec{u}=\frac{2}{n} \vec\Psi_{A_{n-1}}^2$\\
&&&\\\hline
\end{tabular}
\end{center}
\caption{Type, $\vec{s}$, and $C_2 (g^{(i)}) /{\vec\Psi_{g^{(i)}}^2}$
corresponding to the SSSG models of type~B 
associated
to $G$ classical.}
\label{table4}
\end{table}

\ \indent

\vspace{1cm}
\begin{table}[h]
\begin{center}
\begin{tabular}{|c|c|c|c|}\hline
&&&\\
$G/{G_0}$&Type&$\vec{s}$&$
\tilde{C}_2 (g^{(i)})=
{C_2} (g^{(i)}) /{\vec{\Psi}_{g^{(i)}}^2}$\\
&&&\\
\hline\hline
&&&\\
${E_6}/{Sp(4)}$&A2&
$(0,0,0,0,1)$& 6\\
&&&\\
\hline
&&&\\
${E_6}/{F_4}$&A2&
$(1,0,0,0,0)$& 6\\
&&&\\
\hline
&&&\\
${E_7}/{SU(8)}$&A1&
$(0,0,0,0,0,0,1)$& 9\\
&&&\\
\hline
&&&\\
${E_8}/{SO(16)}$&A1&
$(0,0,0,0,0,0,1,0)$& 15\\
&&&\\
\hline
&&&\\
${F_4}/{SO(9)}$&A1&$(0,0,0,0,1)$&9/2\\
&&&\\
\hline
&&&\\
${E_6}/{SU(6) \times SU(2)}$&A1&
$(0,0,0,0,0,0,1)$& 
$\tilde{C}_2({A_5})=21/4$,\,\,$\tilde{C}_2({A_1})=3/4$\\
&&&\\
\hline
&&&\\
${E_7}/{SO(12) \times SU(2)}$&A1&
$(0,0,0,0,0,1,0,0)$& 
$\tilde{C}_2({D_6})=33/4$,\,\,$\tilde{C}_2({A_1})=3/4$\\
&&&\\
\hline
&&&\\
${E_8}/{E_7 \times SU(2)}$&A1&
$(0,1,0,0,0,0,0,0,0)$& 
$\tilde{C}_2({E_7})=57/4$,\,\,$\tilde{C}_2({A_1})=3/4$\\
&&&\\
\hline
&&&\\
${F_4}/{Sp(3) \times SU(2)}$&A1&
$(0,1,0,0,0)$& $\tilde{C}_2({C_3})=15/4$,\,\,
$\tilde{C}_2({A_1})=3/4$\\
&&&\\
\hline
&&&\\
${G_2}/{SU(2) \times SU(2)}$&A1&
$(0,1,0)$&
$\tilde{C}_2(g^{(1)})=3/4$,\,\,$\tilde{C}_2(g^{(2)})=15/4$\\
&&&\\
\hline
&&&\\
${E_6}/{SO(10)  \times U(1)}$&B&
$(1,0,0,0,0,1,0)$& $\tilde{C}_2({D_5})=767/128$,\,\,
$\vec{\Lambda}\cdot\vec{u}={\vec{\Psi}_{D_5}^2}/128$ \\
&&&\\
\hline
&&&\\
${E_7}/{E_6  \times U(1)}$&B&
$(1,0,0,0,0,0,1,0)$& $\tilde{C}_2({E_6})=647/72$,\,\,
$\vec{\Lambda}\cdot\vec{u}={\vec{\Psi}_{E_6}^2}/72$ \\
&&&\\
\hline
\end{tabular}
\end{center}
\caption{ Type, $\vec{s}$ and
${C_2} (g^{(i)}) /{\vec{\Psi}_{g^{(i)}}^2}$ corresponding
to the SSSG models
associated to $G$ exceptional.}
\label{table5}
\end{table}


\vskip 1truecm






\begin{thebibliography}{99}

\bibitem{nt} A.N. Leznov and M.V. Saveliev, Comm. Math.
Phys. 89 (1983) 59;\\ 
J.L.~Miramontes, Nucl. Phys. B547 (1999) 623.

\bibitem{Luiz} L.A. Ferreira, J.L. Miramontes and 
J. S\' anchez Guill\' en, Nucl. Phys. B449 (1995) 631.

\bibitem{ntft} C.R. Fern\' andez--Pousa, M.V. Gallas,
T.J. Hollowood and J.L. Miramontes, Nucl. Phys. B484 (1997)
609.

\bibitem{ls} T.J. Hollowood, J.L. Miramontes and Q-Han Park,
Nucl. Phys. B445 (1995) 451.

\bibitem{park} Q-H. Park, Phys. Lett. B328 (1994) 329;\\
Q-H. Park and H.J.~Shin, Phys. Lett. B359 (1995) 125.

\bibitem{sssg3} I. Bakas, Q-H. Park and Hyun-Jong Shin,
Phys. Lett. B372 (1996) 45.

\bibitem{illToda} T.J.~Hollowood, Nucl. Phys. B384 (1992) 523;\\
S.P.~Khastgir and R.~Sasaki, Prog. Theor. Phys. 95 (1996) 485;\\ 
G.~Tak\'acs and G. Watts, Nucl. Phys. B547 (1999) 538.

\bibitem{CSGBAK} I.~Bakas, Int. J.~Mod. Phys. A9 (1994) 3443.

\bibitem{paras} V. A. Fateev, Int. J. Mod. Phys. A6 (1991)
2109.

\bibitem{CSGMat} R.~K\"oberle and J.A.~Swieca, Phys. Lett. B86 (1979) 209;\\
N.~Dorey and T.J.~Hollowood, Nucl. Phys. B440 (1995)
215.

\bibitem{hsg1}  C.R. Fern\' andez--Pousa, M.V.~Gallas,
T.J.~Hollowood and J.L.~Miramontes, Nucl. Phys. B499 (1997) 673. 

\bibitem{hsg2}
C.R. Fern\' andez-Pousa and J.L.~Miramontes, Nucl. Phys. B518 (1998) 745.

\bibitem{hsg3}
J.L. Miramontes and C.R.~Fern\' andez-Pousa, Phys. Lett. B472 (2000) 392.

\bibitem{TBA} O.A. Castro-Alvaredo, A.~Fring, C.~Korff and J.L.~Miramontes,
{\em Thermodynamic Bethe Ansatz of the Homogeneous Sine-Gordon models\/},
 hep-th/9912196. 

\bibitem{Andreas} A. Fring and C. Korff, {\em Colour valued Scattering
Matrices\/}, hep-th/0001128.

\bibitem{sssg1}
R.D'Auria, T. Regge, and S. Sciuto, Phys. Lett.  89B (1980) 363; Nucl.
Phys. 171B (1980) 167;\\
H.~Eichenherr and K.~Polmeyer, Phys. Lett. 89B (1979) 76;\\ 
H.~Eichenherr, Phys. Lett. 90B (1980) 121;\\ 
V.E.~Zakharov and A.V.~Mikhailov, Sov. Phys. JETP 47 (1978) 1017.

\bibitem{brazhnikov}
V.A. Brazhnikov, Nucl. Phys. B501 (1997) 685.

\bibitem{helgas} S. Helgason, {\em Differential Geometry, 
Lie groups and symmetric spaces}, Academic Press (1990).

\bibitem{parke}
S. Parke, Nucl. Phys. B174 (1980) 166.

\bibitem{Emparan} R. Emparan and I. Sachs, Phys. Rev. Lett. 81 (1998) 2408.

\bibitem{dyon} B.~Julia and A.~Zee, Phys. Rev. D11 (1975) 2227;\\
E.~Tomboulis and G.~Woo, Nucl. Phys. B107 (1976) 221.

\bibitem{twoloop} H.~Aratyn, L.A.~Ferreira, J.F.~Gomes and A.H.~Zimerman,
Phys. Lett. B254 (1991) 372

\bibitem{Gomes} J.F.~Gomes, E.P.~Guevoughlanian, G.M.~Sotkov and
A.H.~Zimerman, {\em Torsionless T-selfdual Affine NA Toda Models\/},
hep-th/0002173.

\bibitem{wzw} 
E. Witten, Comm. Math. Phys. 92 (1984) 455; \\
V.G. Knizhnik and A. Zamolodchikov, Nucl. Phys. B247 (1984) 83.

\bibitem{kac}
V.G. Kac, {\em Infinite dimensional Lie algebras\/} ($3^{rd}$ ed.), 
Cambridge University Press (1990).

\bibitem{olive} P. Goddard and D. Olive, Int. Jour. Mod. Phys.
Vol. 1 No. 2 (1986) 303.

\bibitem{drinfel} M.F. de Groot, T.J. Hollowood and J.L. Miramontes, Comm.
Math. Phys. 145 (1992) 57.

\bibitem{pertcft}
A.B. Zamolodchikov, Adv. Stud. Pure. Math. 19 (1989) 641;
Int. J. Mod. Phys. A 3 (1988) 743; JETP Lett. 46 (1987) 160.

\bibitem{cardy} J.L. Cardy,{\em Conformal invariance and statistical 
mechanics}, in {\em Fields, Strings and Critical Phenomena}, ed. E. 
Br\` ezin and J. Zinn-Justin, Les Houches 1988, Session XLIX
(North-Holland, Amsterdam, 1990) p. 169.
 
\bibitem{baisC} F.A. Bais, P. Bouwknegt, M. Surridge and K. Schoutens, Nucl.
Phys. B304 (1998) 348.

\bibitem{Monopole} F.A.~Bais, Phys. Rev.~D18 (1978) 1206;\\
E.J.~Weinberg, Nucl. Phys.~B167 (1980) 500.

\bibitem{Instanton} C.W.~Bernard, N.H.~Christ, A.H.~Guth and E.J.~Weinberg, 
Phys. Rev.~D16 (1977) 2967.

\bibitem{TodaJap} T.~Nakatsu, Nucl. Phys. B356 (1991) 499.


\end{thebibliography}
\end{document}